\documentclass[aps,pre, reprint, amsmath, amssymb, superscriptaddress, nofootinbib]{revtex4-2}
\usepackage{graphicx} 
\usepackage{braket}
\usepackage{tikz}
\usetikzlibrary{quantikz}
\usepackage{svg}

\pdfoutput=1
\usepackage[utf8]{inputenc}
\usepackage{bm}
\usepackage[english]{babel}
\usepackage[T1]{fontenc}
\usepackage{mathrsfs}
\usepackage[retainorgcmds]{IEEEtrantools}
\usepackage{amsmath}
\usepackage{amssymb}
\usepackage{color}
\usepackage{amsfonts}
\usepackage{times,txfonts}
\usepackage{nicefrac}
\usepackage[colorlinks=true,linkcolor=blue,urlcolor=blue,citecolor=blue,pdfusetitle]{hyperref}
\usepackage{xurl}
\usepackage{mathtools}
\usepackage{isotope}
\usepackage{siunitx}
\usepackage{dsfont}
\usepackage[percent]{overpic}
\usepackage[section]{placeins}
\usepackage{float}
\usepackage{smartdiagram}
\usepackage{float}
\graphicspath{{images/}}

\usepackage{braket}
\usepackage{lipsum}
\usepackage{enumerate}
\usepackage{verbatim}
\usepackage{diagbox}

\begin{document}
\setlength{\abovedisplayskip}{4pt}
\setlength{\belowdisplayskip}{4pt}

\title{Quantum Internet in a Nutshell -- Advancing Quantum Communication with Ion Traps}

\author{Janine Hilder}
\email{j.hilder@neqxt.org}
\affiliation{neQxt GmbH, 63906 Erlenbach am Main, Germany}

\author{Sascha Heu{\ss}en}
\affiliation{neQxt GmbH, 50670 Cologne, Germany}

\author{Anke Ginter}
\email{anke.ginter@bdr.de}
\affiliation{Bundesdruckerei GmbH, Kommandantenstra\ss e 18, 10969 Berlin, Germany}

\author{Andreas Wilke}
\email{andreas.wilke@bdr.de}
\affiliation{Bundesdruckerei GmbH, Kommandantenstra\ss e 18, 10969 Berlin, Germany}

\author{Lukas Postler}
\affiliation{neQxt GmbH, 63906 Erlenbach am Main, Germany}

\author{Ulrich Poschinger}
\affiliation{neQxt GmbH, 63906 Erlenbach am Main, Germany}
\affiliation{QUANTUM, Institut f\"{u}r Physik, Universit\"{a}t Mainz, 55128 Mainz, Germany}

\author{Ferdinand Schmidt-Kaler}
\affiliation{neQxt GmbH, 63906 Erlenbach am Main, Germany}
\affiliation{QUANTUM, Institut f\"{u}r Physik, Universit\"{a}t Mainz, 55128 Mainz, Germany}

\author{Wadim Wormsbecher}
\email{wadim.wormsbecher@bdr.de}
\affiliation{Bundesdruckerei GmbH, Kommandantenstra\ss e 18, 10969 Berlin, Germany}

\begin{abstract}
Quantum Internet in a Nutshell (QI-Nutshell) connects the fields of quantum communication and quantum computing by emulating quantum communication protocols on currently available ion-trap quantum computers. We demonstrate emulations of QKD protocols where the individual steps are mapped to physical operations within our hardware platform. This allows us to not only practically execute established protocols such as BB84 or BBM92, but also include cloning attacks by an eavesdropping party, noise sources and side-channel attacks that are generally hard to include in theoretical QKD security proofs. We deliberately inject noise and investigate its effect on quantum communication protocols. We employ numerical simulations in order to study the incorporation of small quantum error correction (QEC) codes into QKD protocols. We find that these codes can help to suppress the noise level and to monitor the noise profile of the channel. This may enable the communicating parties to detect suspicious deviations from expected noise characteristics as a result of potential eavesdropping. This suggests that QEC  may serve as a means of privacy authentication for quantum communication without altering the transmitted quantum information.
\end{abstract}

\maketitle

\section{Introduction}

The rapidly developing field of quantum technologies offers a variety of opportunities for disruptive innovation. Two main pillars of this research field are quantum computing and quantum communication. Quantum computers hold the promise of computational capabilities far beyond classical computers~\cite{feynman1982,shor1997,preskill2018quantum,dalzell2023quantumalgorithmssurveyapplications}. 
Broad interest in quantum computers surged after Shor's algorithm~\cite{Shor} emerged as a potential threat to commonly used encryption methods. Today's available noisy intermediate-scale quantum (NISQ) devices~\cite{preskill2018quantum} are still far from offering the required quantum computational resources required for breaking encryption protocols such as RSA~\cite{bharti2022nisqreview}. Current estimates assume that cryptographically relevant quantum computers will operational until 2040~\cite{Quantum_Computer_Report_BSI_2024}.

Research in quantum communication is concerned with the development of protocols for secure exchange of information between distant parties utilizing quantum effects~\cite{koudia2024physicallayeraspectsquantum, wiesemann2024consolidatedaccessiblesecurityproof}. Among the most prominent approaches are quantum key distribution (QKD) protocols. In QKD, transmitted quantum states are used in conjunction with a classical communication channel and classical post-processing to securely establish a symmetric key for classical encryption~\cite{rusca2024quantumcryptographyoverviewquantum}. While commericial hardware for QKD applications is already available, the transmission distances achievable via fiber-optical connections are limited and concepts for repeater-based networks are still under development~\cite{mariani2025quantumkeydistributioncomplex, horoschenkoff2025demoquandtcarriergradeqkdnetwork, pittaluga2025long}.
Ultimately, quantum computing and quantum communication are expected to converge into the \emph{quantum internet}~\cite{wehner2018quantuminternet,Quantum_Internet_Challenges}, which will further increase the demand for quantum communication protocols ensuring secure data transmission.

In the context of QKD, a gap between theory, hardware implementations and application requirements is one of the reasons why the security of QKD is still scrutinized by security and standardization agencies~\cite{Position_Paper_QKD, NSA_Link_QKD}. At the time of this writing, major questions regarding the security of quantum communication protocols remain unanswered, such as how assumptions on the hardware implementation impact security guarantees~\cite{tupkary2025qkdsecurityproofsdecoystate, BSI_side_channel_attack_study}. Theoretical analyses of the characteristics of actual quantum technological platforms can be challenging, in particular it can become quite cumbersome to assess to which extent theoretical assumptions are actually met by real-life hardware implementations ~\cite{nahar2025imperfectdetectorsadversarialtasks, trefilov2024intensitycorrelationsdecoystatebb84}. Deviations from the model assumptions and side channel attacks can lead to loopholes, which are not covered by idealized security proofs~\cite{BSI_side_channel_attack_study}. Consequently, new methods and tools are required to verify implementations of QKD and other quantum communication protocols~\cite{ISO_Link_Part_1, ISO_Link_Part_2}.

\begin{figure*}[ht]
    \centering
    \includegraphics[width=\textwidth]{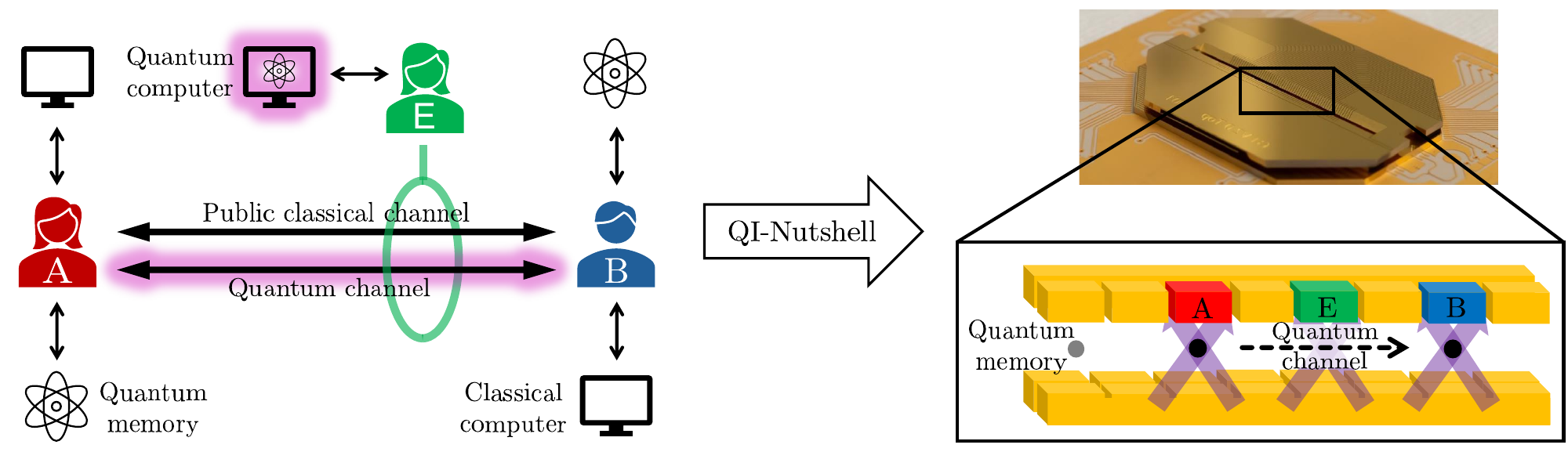}
    \caption{Mapping of quantum communication scenarios to a shuttling-based trapped-ion quantum processor within the QI-Nutshell approach. \textbf{Left:} Two parties, Alice (A) and Bob (B), exchange information via a quantum and a classical channel. As the privacy of a classical channel can not be ensured, the channel is assumed to be a public broadcast. Alice and Bob both have access to a classical computer and to a quantum memory. An eavesdropper Eve (labelled E) can intercept qubits from the quantum channel and has access to a universal quantum computer and a quantum memory. \textbf{Right:} We employ a trapped-ion quantum processor to emulate the scenario described above. Manipulation of the qubits encoded in electronic states of atomic ions, shown as black and gray dots, is achieved by selectively illuminating the ions with laser radiation, shown as purple arrows. A specific quantum communication protocol can be mapped to a sequence of operations acting on the qubits in different areas of the trap, associated to the different communication parties. A quantum channel can be emulated by physically moving qubits between trap sites. Additional storage zones serve for emulating quantum memory, by keeping idle qubits at trap sites which are not accessible via laser radiation. 
	\label{fig:overview}}
\end{figure*}

The development of new security primitives requires systematic evaluation and test procedures that take into account the requirements imposed by the applications. In this work, we propose a framework based on a versatile trapped-ion quantum computer platform for prototyping and testing quantum communication protocols, especially with regard to applicability and security: The \emph{Quantum Internet in a Nutshell} approach, connecting the fields of quantum communication and quantum computing. Since our approach includes the emulation of quantum communication protocols with currently available ion-trap quantum computers, it represents a new use-case for NISQ devices. 

Fig.~\ref{fig:overview} schematically illustrates how the QI-Nutshell approach maps quantum communication networks, i.e.~processing nodes connected via quantum channels, onto an ion-trap quantum processing unit. The key concept - expressing quantum communication protocols via quantum algorithms and executing them on the quantum processor - opens up possibilities for designing, prototyping and characterizing quantum communication networks, protocols and use cases with the aid of a quantum computer. We demonstrate to use the QI-Nutshell approach as a tool to implement penetration-testing methods and security metrics for the transmission of qubits via a quantum channel from a sender Alice (A) to a receiver Bob (B). In particular, we characterize for a prototype QKD scenario to which extent an eavesdropper Eve (E) tapping the quantum channel can compromise the security. 

As one of the currently leading quantum computing hardware platforms,  atomic ions confined in radiofrequency traps provide an excellent match to the QI-Nutshell approach: High operational fidelities and negligible cross-talk errors turn out to be beneficial to investigate noisy communication channels already with few-qubit systems. Different types of noise can be deliberately injected in order to obtain a faithful emulation of quantum communication protocols under realistic conditions, including side channel attacks. Moreover, trapped-ion platforms equipped with the capability to physically shuttle individual ions enable a rather direct mapping of the entities in a quantum communication scenario as illustrated in Fig.~\ref{fig:overview}.

Furthermore, the QI-Nutshell approach can be utilized to investigate how quantum error correction (QEC) procedures can be integrated into quantum communication protocols. We use numerical simulations to show that employing QEC allows us to suppress noise of a quantum communication channel and monitor its characteristics. In particular, the communicating parties can detect suspicious deviations from expected noise characteristics as a consequence of potential eavesdropping. This means that QEC could serve as a fingerprint authentication for quantum communication.

This manuscript is structured as follows: In Sec.~\ref{sec:iontrap} we describe the employed ion-trap quantum processor. Section~\ref{sec:attacks} discusses the emulation of QKD protocols on this processor, and introduces common attack scenarios. Our experimental demonstration of attacks on prototypical QKD scenarios and emulation of realistic noise sources are presented in Sec.~\ref{sec:experiments}. Based on these insights, we explore use cases of QEC codes in more general transmissions of quantum information via numerical simulations in Sec.~\ref{sec:qec}. Our findings are summarized in Sec.~\ref{sec:conclusion}, where we discuss potential applications of the QI-Nutshell approach.

\section{Trapped-ion quantum processor}\label{sec:iontrap}

Hardware platforms based on atomic ions trapped in radio-frequency traps are one of the leading contenders within the rapid evolution of quantum computers. In radio-frequency traps, a combination of static and oscillating electric fields allows for stable confinement of atomic ions, as well as excellent isolation from undesired environmental interactions. The required state preparation and measurement (SPAM) operations as well as quantum gate operations can be realized using laser radiation in conjunction with a rich toolkit of methods from atomic physics. These prerequisites are the cornerstone of the particular strengths of trapped-ion platforms, namely the comparatively high achievable fidelities of all required qubit operations, long coherence times, low cross-talk and possible connectivity beyond nearest-neighbor coupling. Current trapped-ion quantum processing units from small to intermediate size fall into two categories: One approach consists of maintaining all qubit ions within a single confining potential, where they can form a linear chain and can be individually addressed with laser beams~\cite{pogorelov2021compactqc}. The other approach, originally dubbed \emph{quantum CCD}~\cite{kielpinski2002qccd}, relies on micro-structured, multi-electrode ion traps, which allow for simultaneously storing groups of ions in distinct trap potential wells. The quantum register can be dynamically reconfigured by \emph{shuttling operations}~\cite{kaushal2020shuttling}, where ions are moved within the trap by changing the applied electrode voltages. The shuttling-based approach circumvents some of the challenges and operational errors which increase with the size of the ion chain and leads to improved scaling in the intermediate size regime. These developments have recently culminated in the Quantinuum H2 platform being able to handle up to 56 fully functional qubits, displaying yet-unchallenged quantum advantage~\cite{moses2023racetrack, decross2024computationalpowerrandomquantum, certified-randomness-JPMorgan-QNTM}. \\
For this work, the achievable register size is not the key parameter, but rather \emph{cross-talk} errors are of fundamental interest: These are coherent errors occurring on addressing-based platforms, due to residual drive fields coherently interacting with idle 'spectator' qubits. Such errors are not inherently present in real-life quantum communication scenarios, due to the spatial separation of the communicating parties and therefore have to be avoided in the emulation of such protocols. The low error rates of gate and SPAM operations achievable on shuttling-based small-scale quantum processing nodes, in conjunction with the virtually complete suppression of cross-talk errors, renders such platforms to be the ideal playground for versatile and faithful emulation of quantum communication protocols.

\begin{figure}[htp]
    \centering
    \includegraphics[width=\columnwidth]{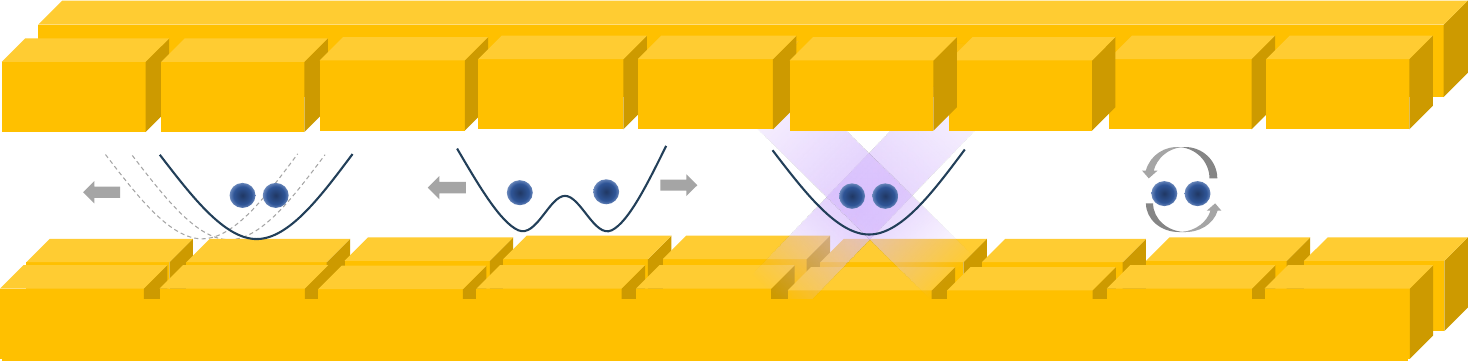}
    \caption{Trapped-ion quantum computing architecture including register reconfiguration operations. The operations from left to right are transport of ion crystal, separation/merge, laser-driven quantum gates and physical ion swap.}
	\label{fig:neQxt_Architecture}
\end{figure}

The experimental results presented in this work were obtained on a trapped-ion quantum computer that consists of a linear, segmented radio-frequency trap with 32 storage segments and one laser interaction zone, where all SPAM and gate operations are driven by laser beams. This setup works with multiple trapping potentials, containing one or two qubit ions each. Effective all-to-all connectivity is achieved by reconfiguration operations such as transport of ions between the memory and processing regions, separation and recombination of ion crystals and position exchange within a crystal as illustrated in Fig.~\ref{fig:neQxt_Architecture}. The gate fidelities determined with the help of randomized and cycle benchmarking are 99.98(1)\% for single-qubit gates and 99.6(2)\% for two-qubit gates \cite{hilder2022fault}. Furthermore, SPAM error rates of less than 0.1\% are achieved. With this system, entanglement of up to 6 qubits was successfully demonstrated and one of the first realizations of a shuttling-based fault-tolerant parity measurement with four data and two auxiliary qubits was demonstrated~\cite{hilder2022fault}.

The control software for the neQxt quantum computer is separated into a high-level and a low-level stack as well as additional logging and calibration frameworks. Multiple compilation stages are used to translate a quantum algorithm specified by a user into hardware commands for execution. First, a hardware-agnostic framework such as Qiskit stores the input circuit in OpenQASM format~\cite{cross2017openquantumassemblylanguage}, which is then transpiled into a quantum circuit using gates and operations native to the hardware~\cite{Kreppel2023quantumcircuit}. This is followed by a shuttling compiler, solving the qubit-to-ion mapping and routing problem using an efficient heuristic~\cite{Durandau2023automatedgeneration}. The low-level stack sequencer assembles all real-time operations and translates them into sequences consisting of time-varying electrode voltages, radio-frequency pulses and trigger pulses, which are sent to the hardware drivers for hardware-level processing. All processing layers within the high- and low-level stacks are fully automated.

\section{Emulating QKD}\label{sec:attacks}

In this section, we explain the explicit mapping procedure of the QKD protocols BB84 and BBM92 onto the ion-trap architecture according to our QI-Nutshell approach. We include an eavesdropper who has access to a quantum register and executes cloning routines on the distributed quantum state. Throughout this work, we use \textit{emulate} when QI-Nutshell replicates a communication protocol on a trapped-ion quantum computer, and \textit{simulate} when the same protocol is executed on classical hardware for comparison, e.g.~using the Qiskit Aer simulator backend.

\subsection{Brief introduction to QKD}

QKD describes a collection of protocols for symmetric secure sharing of cryptographic keys between two parties (commonly called Alice and Bob) who wish to use the key to encrypt messages for secure classical communication. QKD protocols are typically classified into two main categories, namely continuous-variable (CV-) and discrete-variable (DV-) QKD. Both describe the distribution of quantum information through a quantum channel between Alice and Bob. Their difference lies in the nature of the degrees of freedom that carry the quantum information. In CV-QKD the quantum information is encoded in continuous degrees of freedom, e.g. in the amplitude and phase quadratures of electromagnetic field modes. By contrast, DV-QKD uses discrete degrees of freedom, e.g. the polarization of photons. For a comprehensive overview on technological implementations, the reader is referred to~\cite{BSI_side_channel_attack_study}. Throughout this work, we focus exclusively on two common DV-QKD protocols: The BB84 protocol~\cite{BEN2014} and the BBM92 protocol~\cite{bennett1992quantum}. Derived variants of both protocols are used in current technological implementations of QKD~\cite{PAMQKD}. BB84 is a prepare-and-measure protocol, i.e. Alice prepares a qubit and sends it to Bob, who measures it upon reception. The BBM92 protocol is an entanglement-based protocol, where pairs of entangled qubits are shared between both parties, who subsequently measure their respective qubits~\cite{NiedersachsenQKD2025}.

In general, QKD protocols require a quantum channel for the distribution of quantum information and a classical, authenticated, but fully public channel between Alice and Bob. After transmission and measurement of the quantum information, Alice and Bob are said to possess their respective raw keys. Due to intrinsic noise or an active eavesdropper (commonly called Eve) tapping the quantum channel, the raw keys will generally differ between Alice and Bob. To obtain an error-free key, Alice and Bob execute an error reconciliation strategy using their classical channel to exchange data derived from their raw keys. After eliminating all errors, they apply the final step of privacy amplification which eliminates any remaining correlations with Eve, so that they end up with the final, secure key~\cite{tupkary2025qkdsecurityproofsdecoystate, renner2006securityquantumkeydistribution, wiesemann2024consolidatedaccessiblesecurityproof}. Depending on the specific protocol, various classical verification steps are performed in between. Generally, security proofs for QKD protocols require a set of assumptions, as well as a choice of attack strategies by Eve. For a recent security proof we refer the reader to Ref.~\cite{wiesemann2024consolidatedaccessiblesecurityproof}. A modern overview of necessary techniques and assumptions is found in Ref.~\cite{tupkary2025qkdsecurityproofsdecoystate}.

\subsection{BB84 and BBM92 protocols including eavesdropping}

In this section we give a brief review of both the BB84 and BBM92 protocols, including an adversarial party who is performing an individual cloning attack~\cite{PhysRevA.65.052310, PhysRevA.61.052304} on the part of the protocols involving transmission of qubits over a quantum channel.

In the following, we assume completely error-free supporting-technological building blocks for all parties, e.g. perfectly uniform random number generators, error-free interfaces between the parties, no exploitable side channels, and error-free quantum gate operations. We further assume that Eve has no access to the local systems of Alice and Bob, but can interact with the quantum channel and has full access to any message that is transmitted through the authenticated classical channel. We will explicitly mention whenever we relax these assumption throughout this analysis.

\subsubsection{BB84}

We use the same convention for the BB84 protocol as in Ref.~\cite{decker2025qkdquantummachinelearning} including an adversary Eve. A raw key of length $N$ results from repeating the following sequence of steps $N$ times:

\begin{enumerate}
    \item Alice randomly selects a classical bit $x_A \in \{0, 1\}$, stores it in a classical memory and prepares a corresponding quantum state $\ket{x_A}$. Note that $\ket{1} = X\,\ket{0}$, where $X$ denotes the Pauli-X gate, and that $\{\ket{0}, \ket{1}\}$ is called Z-basis. Next, Alice randomly selects a basis $b_A \in \{0, 1\}$ and applies either the identity operation to $\ket{x_A}$ for $b_A=0$ or the Hadamard operation $H$ for $b_A=1$, transforming Z-basis states into the X-basis states:  $H \ket{0}=\ket{+},\; H\ket{1}=\ket{-}$, where $\ket{\pm} = \tfrac{1}{\sqrt{2}}\,\left(\ket{0} \pm \ket{1}\right)$.
    \item Alice sends the prepared state to Bob via the quantum channel.
    \item Eve intercepts the state transmission by executing a cloning circuit using the transmitted qubit and a blank qubit. The cloned quantum state is not measured but kept in a quantum memory while Alice's original qubit is passed on towards Bob.
    \item Bob receives the quantum state and randomly selects a basis $b_B \in \{0, 1\}$ and applies the corresponding measurement gate to the qubit, followed by a Z-basis measurement and storage of the classical bit result. If the basis choices of Alice and Bob match, they always obtain the same result in the absence of any noise or eavesdropper in the channel.
    \item Bob communicates his basis choice to Alice using the authenticated classical channel. Eve has access to this information. If the bases of Alice and Bob do not match, the corresponding bit is discarded from classical memory. Accordingly, Eve also discards the respective stored qubit. If the bases match, Eve can apply further gate operations to the quantum state before measuring it and storing the bit in classical memory.
\end{enumerate}

All three parties now possess a raw key. In a real QKD scenario, Alice and Bob would now use the authenticated classical channel to estimate the error rate in their raw keys. The QKD protocol will abort if the error rate between Alice's and Bob's raw keys surpasses a critical quantum bit error rate (QBER). For individual attacks on BB84, this QBER is approximately $14.5\%$~\cite{Bruss2000}, meaning that the fidelity between the raw keys of Alice and Bob must be at least $0.855$. If this is the case, Alice and Bob proceed with an error-reconciliation scheme. For our purpose, we are only interested in Eve's ability to obtain the raw key. BB84 is well-studied and the optimal individual attack, in the absence of any noise, is the phase covariant cloning machine (PCCM)~\cite{Bruss2000}, which we will explore below. As has been shown in Ref.~\cite{decker2025qkdquantummachinelearning}, at least one stronger attack than the PCCM -- called \emph{imbalanced cloner} -- exists in the presence of noise in the quantum communication channel.

\subsubsection{BBM92}

\label{sec:bbm92intro}
The BBM92 protocol~\cite{bennett1992quantum} was developed as a reaction to the E91 protocol~\cite{ekert1991quantum}, in which entanglement is used as a resource. Security is analyzed by performing Bell tests on the distributed qubits and verifying non-locality. However, it was realized quickly that security could also be investigated without performing Bell tests, and consequently E91 was simplified and reformulated as BBM92 allowing for an easier physical implementation.

Again, a raw key of length $N$ is obtained by repeating the following sequence of steps $N$ times:

\begin{enumerate}
    \item A source generates a pair of qubits in the Bell state $\ket{\Phi^+} = \frac{1}{\sqrt{2}}\, \left(\ket{00} + \ket{11}\right)$, which are subsequently distributed such that one qubit is sent to Alice and the other qubit to Bob. Note that there are now two quantum channels that can be attacked.
    \item Eve intercepts, for example, the state transmission between the source and Bob and clones the state in the same way as previously explained for BB84.
    \item Alice and Bob independently and randomly choose bases $b_A, b_B \in \{0, 1\}$, apply the respective measurement gates, measure in the Z-basis and store the classical results in their respective classical memories.
    \item Alice and Bob publicly communicate their basis choices and discard the bit if measured with differing bases. Just as for BB84, Eve may perform additional operations on her qubit before measuring as well.
\end{enumerate}

Analogous to BB84, in BBM92 the raw key is post-processed. While BBM92 is conceptually equivalent to BB84, the protocols differ in the attacking possibilities. If we imagine that the source is positioned right next to Alice's location, Bob would not be able to distinguish whether his qubit was prepared by Alice or the source. In this case, BBM92 would be the same as BB84 from Bob's perspective. However, using an external source decreases the complexity for Alice's node, which entails fewer possibilities for side-channel attacks~\cite{source_replacement}. Remarkably, it can be shown that for BBM92, even if Eve controls the source, security of the protocol is still maintained~\cite{bennett1992quantum}. 

\subsection{Emulating QKD protocols including attacks on trapped-ion platforms}

Both, BB84 and BBM92, can be straightforwardly expressed through quantum circuits which can be executed on gate-based quantum computers. To understand how the hardware resources of a trapped-ion platform can be used to model QKD scenarios, it is important to list the entities of which such a scenario is necessarily comprised:

\begin{itemize}
\item
A number of \textbf{parties} participating in the protocol, here, a sender Alice, a receiver Bob and an eavesdropper Eve. The parties are henceforth abbreviated as A, B and E. Each party operates a quantum information processing unit (QPU), offering capabilities to prepare and manipulate qubits via reset and gate operations and to measure qubits in a specific basis.
\item
At least one \textbf{quantum channel} for transmitting qubits between parties. Typically, A sends qubits to B via a one-way channel. E can intercept qubits on this channel and resend qubits she has manipulated to B.
\item
An authenticated \textbf{classical communication channel}. As classical communication is assumed to be inherently unsafe, one typically assumes that A and B communicate via unencrypted broadcasts, such that all information communicated is also available to E.
\end{itemize}
First, it is important to realize that, given typical physical distances of trapped-ion qubits (few \si{\micro\meter} to few \si{\milli\meter}) and typical operation timescales (few tens of \si{\micro\second} for gate operations to few \si{\milli\second} for measurements), space-like separations of measurement events cannot be realized on these platforms. Yet it is possible to associate different communication parties with different sites in a trap architecture. The experimental setup used within this work features only one processing zone. Thus different entities participating in a protocol have to be associated to different parts of a sequence of operations, which can be seen as different controllers acting on the qubits at different times. 

\begin{figure}[!htp]
    \centering
    \includegraphics[width=\columnwidth,trim={2cm 2cm 14cm 2cm},clip]{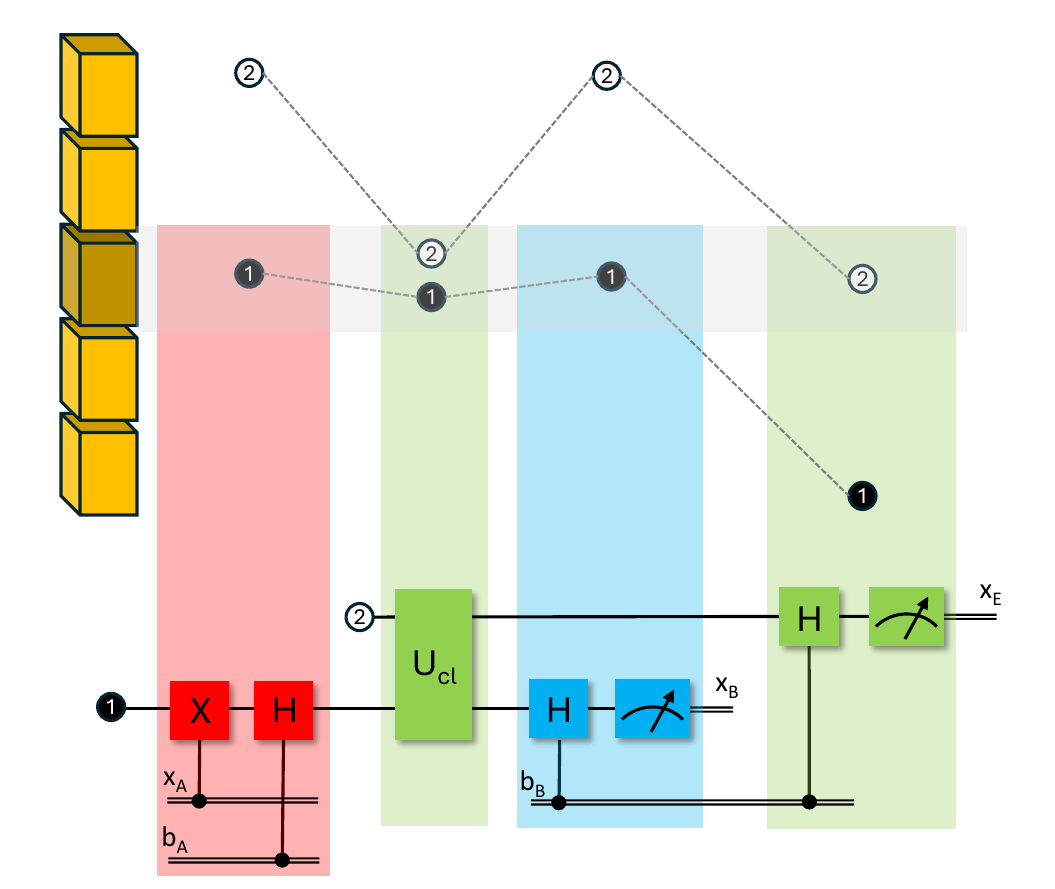}
    \caption{Emulation of an attack on BB84 on a trapped-ion QPU with a single processing zone. Shown are some storage segments of the trap (yellow) including the processing zone (dark yellow). The colored boxes indicate which party assumes control over the processing zone in order to execute parts of the circuit (bottom): Alice (red), Bob (blue) and Eve (green). The input bits are Alice's data bit $x_A$ and measurement basis $b_A$, as well as Bob's measurement basis $b_B$, which is also available to Eve as it is broadcast at a later stage. The output is given by Bob's and Eve's measurement results $x_B$ and $x_E$.}
	\label{fig:BB84attack}
\end{figure}

Figure~\ref{fig:BB84attack} illustrates this concept on the basis of BB84 including an eavesdropper using two trapped-ion qubits which are manipulated in a single processing zone, as explained in the following: First, A (red) possesses control and prepares a qubit based on the random bit value to be set, $x_A$, and the random encoding basis $b_A$. Then, E (green) takes over, intercepts the qubit and uses an additional blank qubit and a cloning machine to create an imperfect copy of the original state. One of the qubits is routed to B (blue), who then assumes control of the processing zone and measures in the random basis $b_B$. Finally, manipulations and a measurement is carried out by E (green) on the remaining qubit, using E's available information about the chosen preparation and measurement bases. Upon the final post-processing, the data which would be broadcast in a real-life realization is considered as being available to all parties. This way of mapping QKD scenarios to a trapped-ion QPU is conceptually valid, the differences being that space-like detection events cannot be realized and the noise is different from actual real-life devices for quantum communication. Possible errors from transmission along a noisy quantum channel, preparation and readout errors exceeding the native error rates of the platform, as well as side-channel attacks, can be emulated by appropriate means.

The native noise in trapped-ion quantum computing platforms is of rather different type as compared to the noise occurring in real-life QKD scenarios, where photonic qubits are transmitted via optical fibers and detected on single-photon detectors. Given the long coherence times and excellent gate and SPAM fidelities of trapped-ion qubits, the primary challenge is to introduce tailored noise in order to achieve a realistic emulation of photonic QKD setups. First, we briefly discuss the relevant native error sources of trapped-ion QPUs. Given that the circuits used for realizing simple QKD protocols require only a few qubits and low circuit depth, native decoherence rates and gate errors will have minor impact. While typical timescales for gate, shuttling and readout operations are in the range of tens of \si{\micro\second}, decoherence timescales on the order of seconds or even longer can be achieved with trapped-ion qubits. Moreover, cutting-edge trapped-ion platforms achieve gate and readout error rates on the order of $10^{-3}$ per operation. By contrast, QKD setups mainly suffer from photon loss upon transmission and limited quantum efficiency of single-photon detectors. The best single photon detectors currently available attain a detection efficiency of about 98\% at telecom wavelengths~\cite{REDDY2020}. Aiming at realistic emulation of a QKD scenario, such errors need to be injected. A simple possibility for this would be a probabilistic post-processing stage, where the readout statistics are modified to model qubit loss. On the physical level, such errors can be injected by controlled depletion of population from the qubit subspace to additional (meta)stable states after protocol stages pertaining to qubit transmission or prior to readout.

While the employed architecture does not affect fundamental concepts and conclusions, one might think about realizing emulations of attacks on QKD protocols on trapped-ion quantum computing architectures featuring multiple processing zones. This yields a higher degree of correspondence to real-life QKD settings as compared to the previously discussed approach, as the different entities participating in the protocol of interest are spatially separated. The main physical difference is merely that the involved quantum channels are associated with actual physical transport through a segmented ion trap, and additional error sources from these operations need to be taken into account.


\section{Measurement Results}
\label{sec:experiments}

\begin{figure*}[tp!]
    \centering
    \includegraphics[width=0.75\textwidth]{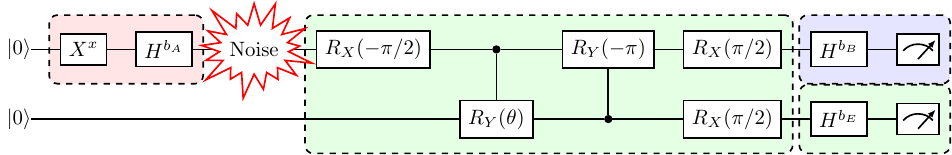}
    \caption{Circuit for the BB84 protocol including optional noise in the channel between A (red) and B (blue). A PCCM attack by E (green) is performed. State preparation is implemented by A with optional $X$ and $H$ gates depending on choice of bit and basis configuration. Final $H$ rotations by B and E similarly depend on the basis choice. We consider independent X and Z errors as the noise channel or other custom noise implementations (see Sec.~\ref{sec:attacks}).}
	\label{fig:BB84_circ_meas}
\end{figure*}

This section shows how the trapped-ion QPU described in Sec.~\ref{sec:iontrap} can be used within the QI-Nutshell framework to investigate the behavior of different QKD protocols in the presence of attacks and tailored noise. The emulation of the qubit transmission channel within this approach allows one to execute attack protocols directly at the quantum level. We verify known results from simulations found in Ref.~\cite{decker2025qkdquantummachinelearning} for the BB84 protocol and employ a quantum machine learning method to retrieve an optimal cloning circuit. Moreover, we demonstrate an attack on the BBM92 protocol described in Sec.~\ref{sec:bbm92intro}. We also show physical realizations of building blocks for the emulation of side-channel attacks.

\subsection{Emulating attacks on QKD Protocols}

\begin{figure}[htp!]
    \centering
    \includegraphics[width=0.8\columnwidth]{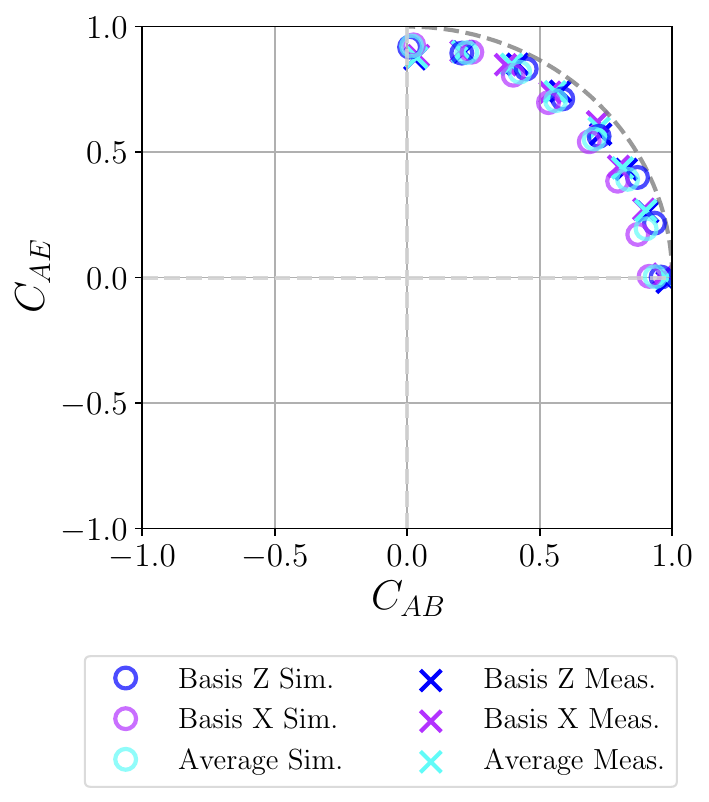}
    \caption{Correlations $C_{AB}$ and $C_{AE}$ in case of a PCCM applied to the BB84 protocol without any additionally injected errors. The corresponding quantum circuit is shown in Fig.~\ref{fig:BB84_circ_meas}. Dashed lines are the expected values from analytical calculations. The cross markers show measurement results. Circle markers are results of numerical simulations with circuit-level depolarizing noise of strength $1\%$ on all gates (see Sec.~\ref{sec:qec}). Experiments where the chosen basis is $X$ ($Z$) are shown in purple (blue). The average over those two cases is shown in cyan. Each data point is the average over 2000 circuit executions (1000 for each bit configuration) so that the statistical error from projection noise is 0.022 in the worst case at $C_{AB} = 0$ or $C_{AE}$ = 0. Numerical simulations coincide well with the measured data. Error bars are not shown for clarity.}
	\label{fig:BB84_PCCM_no_err}
\end{figure}

\subsubsection{BB84}

\begin{figure}[htp!]
    \centering
    \includegraphics[width=\columnwidth]{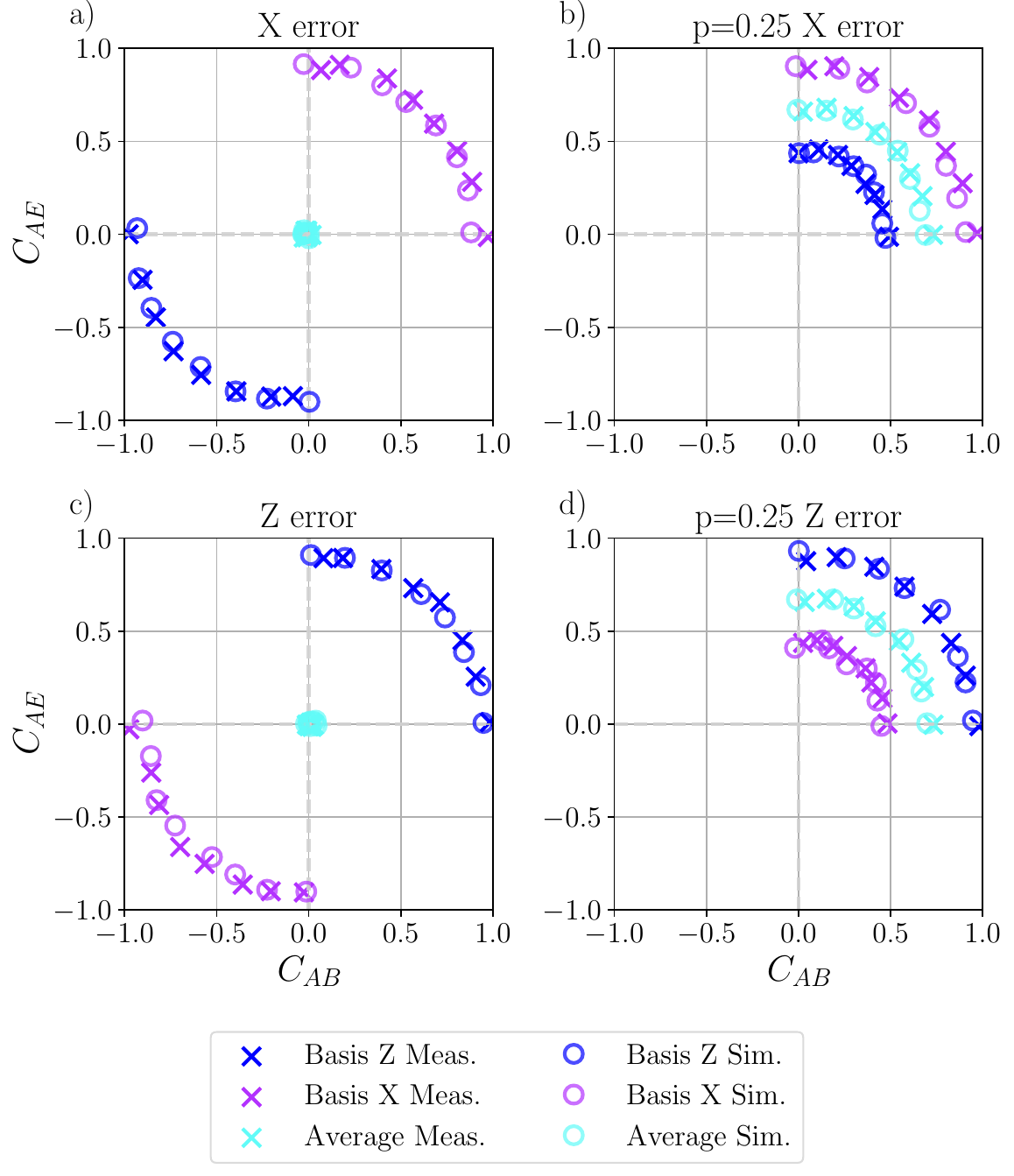}
    \caption{Correlations $C_{AB}$ and $C_{AE}$ in case of a PCCM applied to the BB84 protocol with additionally injected errors. In \textbf{a)} and \textbf{b)} an $X$ error is injected deterministically and with a probability of 0.25, respectively. Subfigures \textbf{c)} and \textbf{d)} show results for the analogous case of injected $Z$ errors. The corresponding quantum circuit is shown in Fig.~\ref{fig:BB84_circ_meas}. The cross markers show measurement results. Circle markers are results of numerical simulations with circuit-level depolarizing noise of strength $1\%$ on all gates (see Eqs.~\eqref{eq:depol} and \eqref{eq:depol_circ}). Experiments where the chosen basis is $X$ ($Z$) are shown in blue (purple). The average over those two cases is shown in cyan. Each data point is the average over 2000 circuit executions (1000 for each bit configuration) so that the statistical error from projection noise is 0.022 in the worst case at $C_{AB} = 0$ or $C_{AE}$ = 0. Numerical simulations coincide well with the measured data. Error bars are not shown for clarity.}
	\label{fig:BB84_all_x}
\end{figure}

The optimal individual attack on BB84 in the case of an error-free channel is given by a PCCM~\cite{Bruss2000}. We first emulate the functionality of BB84 under PCCM attack on the trapped-ion QPU, executing the protocol shown in Fig.~\ref{fig:BB84attack} with the complete circuit including the cloner shown in Fig.~\ref{fig:BB84_circ_meas}. The success rate for the transmission between A and B is quantified by the state fidelity $F_{AB}$ between A's initially prepared qubit and the qubit finally routed to B, averaged over the preparation and measurement bases and the binary input state, and postselected for equal measurement bases (referred to as sifting in a QKD context). Likewise, the success rate for eavesdropping by E is quantified by the state fidelity $F_{AE}$ between A's initial state and the state of E's additional qubit after execution of the PCCM, also averaged and sifted. For the PCCM circuit shown in Fig.~\ref{fig:BB84_circ_meas}, without additional noise, these average fidelities are

\begin{equation}
    \begin{split}
    F_{AB} = \frac{1+\cos\theta/2}{2} \\
    F_{AE} = \frac{1+\sin\theta/2}{2},
    \end{split}
\end{equation}

where the attack angle $\theta$ expresses how much information is cloned from the intercepted qubit onto E's qubit. In particular, $\theta=0$ corresponds to no attack, while $\theta=\pi/2$ corresponds to the symmetric case $F_{AB}=F_{AE}\approx 0.85$. The average fidelities are related to the correlations between A's initial bit values $x_A$ and B's binary measurement outcomes $x_{B}$ as

\begin{equation}
    C_{AB} = \overline{(2x_A-1)(2x_B-1)} = 2F_{AB} -1 \label{eq:centered_fid}
\end{equation}

and likewise for $C_{AE}$. The overline corresponds to averaging over independent protocol runs and sifting. These correlations can be estimated from experimental data, by carrying out a finite number of independent runs of the protocol on the trapped-ion QPU thus generating an emulated secret key.

\begin{figure*}[!htp]
    \centering
    \includegraphics[width=0.99\textwidth]{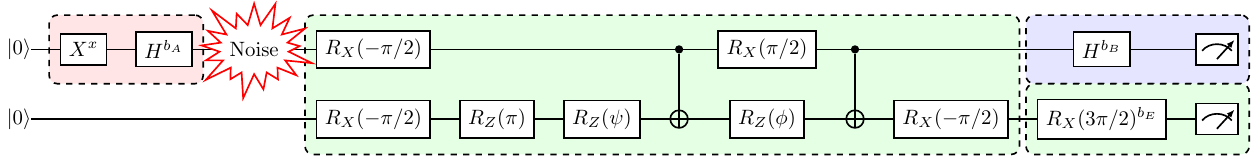}
    \caption{Circuit for the BB84 protocol including optional noise in the channel between A (red) and B (blue). An imbalanced cloning attack by Eve (green) is performed. The parameters $\psi$ and $\phi$ are chosen according to Eq.~\eqref{eqn:imbalanced_cloner_phi} such that the correlations between the prepared bit value and the measurement outcomes for Bob and Eve are maximized under an imbalanced choice of preparation bases.}
	\label{fig:BB84_ASYM_circ}
\end{figure*}

Figure~\ref{fig:BB84_PCCM_no_err} shows measured estimates of correlation values resulting from 1000 independent runs of the protocol without injected errors for each bit and basis configuration, leading to 2000 runs for each basis and 4000 runs for the average. Shown are the results averaging over measurement bases and additional subdivision by the particular measurement bases $X$ or $Z$. Data is taken for eight different attack angles $\theta$, varying between 0 and $\pi$. As can be seen, the correlation of the results can be gradually transferred from A/B to A/E via the attack angle, irrespective of the choice of the measurement basis. In compliance with the no-cloning theorem, the ideal values of $C_{AB}^2+C_{AE}^2$ lie on the unit circle, as confirmed by statevector simulations. The measured correlations fall short of the expected values by up to 12.9\% (7.3\% on average), which is mainly attributed to two-qubit gate errors.

\begin{figure*}[!htb]
    \centering
    \includegraphics[width=0.7\textwidth]{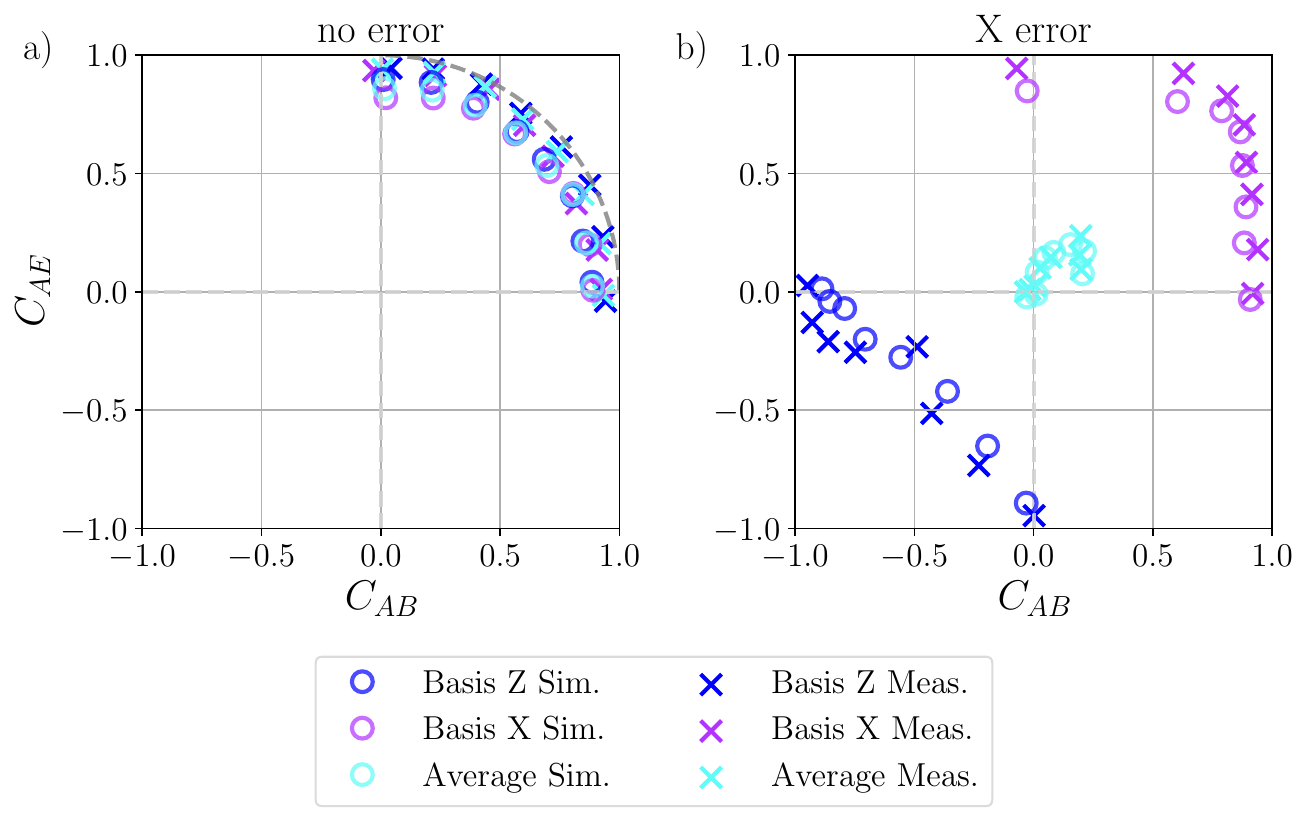}
    \caption{Correlations $C_{AB}$ and $C_{AE}$ in case of an imbalanced cloner attack applied to the BB84 protocol. The employed circuit is shown in Fig.~\ref{fig:BB84_ASYM_circ}. \textbf{a)} Measured correlation estimates for a symmetric, error free channel. With the optimum tuning angle $\phi$ computed from Eq.~\eqref{eqn:imbalanced_cloner_phi} and error rate $p=0$, we retrieve the behavior of a symmetric cloner, showing performance identical to the PCCM results displayed in Fig.~\ref{fig:BB84_PCCM_no_err}. \textbf{b)} Measured correlation estimates for an imbalanced cloner with tuning angle $\phi$ in the circuit chosen for X errors occurring at a finite rate of $p=0.25$. However, the data shown is obtained for the extreme case of deterministically applied $X$ errors. A number of 1000 identical runs were used for each bit and basis configuration. Most of the experimentally measured data points (crosses) lie closer to the ideally-expected values (dashed lines) than data points from numerical simulations (circles) with circuit-level depolarizing noise of strength $1\%$ on all gates (see Sec.~\ref{sec:qec}).}
\label{fig:BB84_ASYM_meas}
\end{figure*}

A key feature of the QI-Nutshell approach is the possibility to deliberately inject errors at various stages of the protocol. Here, we emulate errors occurring at A's node or up until the qubit is intercepted by E, by injecting Pauli $X$ or $Z$ errors before the cloning unitary is carried out. Figure~\ref{fig:BB84_all_x} shows measured correlations $C_{AB}$ and $C_{AE}$, again for different attack angles $\theta$ between 0 and $\pi$, and for the cases of either deterministic Pauli errors or Pauli errors randomly injected at finite rates of $p=0.25$. It can be seen that, as expected, $X$($Z$) errors flip the sign of the correlations of the $Z$($X$) basis measurements, leading to vanishing overall correlation for both B and E, with respect to the original bit value to be transmitted. The more realistic case of finite error rates derives from these results. The data shown in Fig.~\ref{fig:BB84_all_x} for the -- deliberately exaggerated -- case of $p=0.25$ separately for each Pauli error is obtained by combination of results from the error-free data shown in Fig.~\ref{fig:BB84_PCCM_no_err} and the data with deterministic errors in Fig.~\ref{fig:BB84_all_x}. It can be seen that the effect of the errors on the correlations becomes asymmetric, and that residual overall correlations remain upon averaging over the bases.

Next, we use an \emph{imbalanced cloner} \cite{decker2025qkdquantummachinelearning}: E employs a cloning unitary which provides better cloning fidelity in the $X$ basis, at the expense of reduced cloning fidelity in the $Z$ basis. The corresponding circuit is shown in Fig.~\ref{fig:BB84_ASYM_circ}, which is parameterized by two angles $\psi$ and $\phi$. While $\psi$ now takes the role of the attack angle, the additional parameter $\phi$ now allows for tuning the overall cloning fidelity between the $X$ and $Z$ bases. In case of asymmetric channel noise, such a cloner could outperform a PCCM given that the choice of $\phi$ is matched to $\psi$ and the basis asymmetry. Considering highly asymmetric errors, i.e. only $X$ errors in the quantum channel between A and B, occurring at a rate $p$, we can express the fidelity $\Tilde{F}_{AB}^{(i)}$ for the $Z$ basis states $i \in \{0,1\}$ as
\begin{align}
    \Tilde{F}_{AB}^{(i)}  = (1-p) F_{AB}^{(i)} + p \left(1-F_{AB}^{(\neg 1)}\right)
\end{align}
in terms of the bare state fidelity $F_{AB}^{(i)}$ of B's received qubit with respect to A's prepared Z basis state $\ket{i}$, which is affected by E's cloning procedure, but not by additional channel noise. We calculate the average fidelity in the $Z$ basis $\Tilde{F}_{AB, Z} = \left(\Tilde{F}_{AB}^{(0)} + \Tilde{F}_{AB}^{(1)}\right)/2$ and use Eq.~\eqref{eq:centered_fid} to obtain
\begin{align}
    \Tilde{C}_{AB, Z} &= (1 - 2p) \, C_{AB, Z}~~~\text{and}~~~\Tilde{C}_{AE, Z} = (1 - 2p) \, C_{AE, Z},
\end{align}
respectively. The noise-affected correlations in the $X$ basis $\Tilde{C}_{AB, X}$ and $\Tilde{C}_{AE, X}$ remain unchanged in case that only $X$ errors are considered. From the calculation given in Ref.~\cite{decker2025qkdquantummachinelearning}, we  obtain the optimal tuning angle
\begin{align}
    \phi =& -\arctan\left((1-2p)^2\cot\psi\right).\label{eqn:imbalanced_cloner_phi}
\end{align}
Figure~\ref{fig:BB84_ASYM_meas} shows measured correlation estimates, again for different attack angles between $\psi=0$ and $\psi=\pi/2$ (note that for the imbalanced cloner, the maximum attack angle is $\pi/2$, while for the PCCM it is $\pi$)). As a sanity check, we test the case without asymmetry and retrieve the performance of the PCCM as shown in Fig.~\ref{fig:BB84_PCCM_no_err}. Then, we study the extreme case of attack angles chosen according to $p=0.25$ and deterministic bit flip in order to compare to Fig.~\ref{fig:BB84_all_x} a). It can be seen that upon postselection on the X-basis, the joint correlations $C_{AB}$ and $C_{AE}$ are beyond the intrinsic limitations of the PCCM. Furthermore, in contrast to the behavior of the PCCM for deterministic flips, a residual joint correlation remains upon averaging over the basis choices at equal rates.

\subsubsection{Quantum circuit learning}

The QI-Nutshell framework does not only suit as a testbed for existing QKD attack protocols. It can also be extended to investigate new approaches such as quantum machine learning (QML). It has been shown in Ref.~\cite{decker2025qkdquantummachinelearning} that quantum circuit learning (QCL), a subdiscipline of QML, can be used to find optimal attacks on the BB84 protocol. Here, we experimentally demonstrate such a QCL attack on BB84, using a hybrid quantum-classical algorithm where the weights of the parameterized circuit are classically optimized and updated by minimizing a loss function. The loss function is defined as

\begin{equation}
    \mathcal{L}(\theta) = \alpha(F_{AB}(\theta)-f)^2-F_{AE}(\theta),
\end{equation}

\begin{figure}[!htp]
    \centering
    \includegraphics[width=0.8\columnwidth]{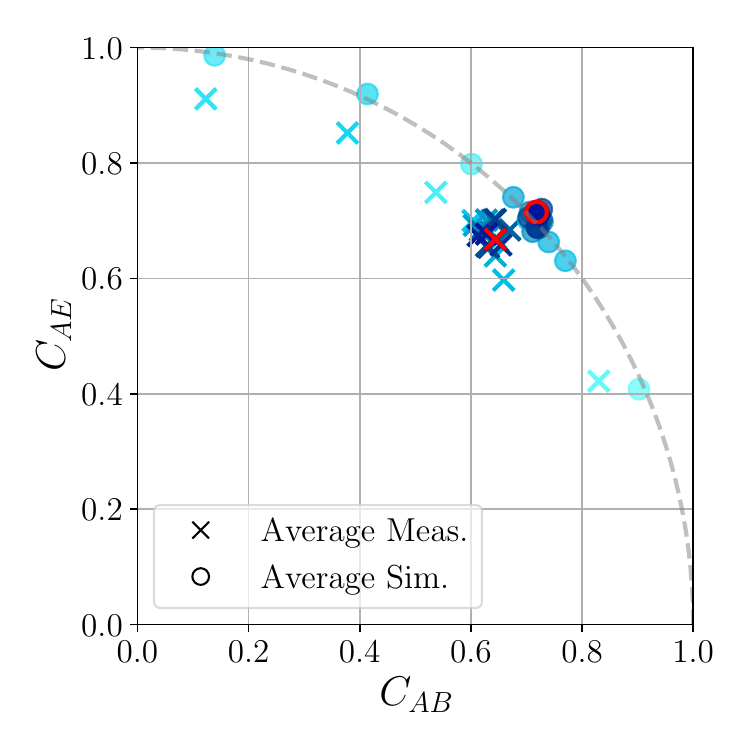}
    \caption{Demonstration of a hybrid quantum-classical approach using a PCCM attack. The attack angle is trained to achieve a specified fidelity $F_{AB}=f$, while simultaneously maximizing $F_{AE}$. Evaluations of the loss function during the optimization iterations are shown in light blue to dark blue with increasing iteration, in red the final result. A number of 500 identical runs were used for each of the four bit and basis configurations. The optimization converged after 20 iterations. Initial parameter was chosen randomly. The simulation is based on $\theta$ values of each iteration evaluated using the Qiskit Aer simulator.}
	\label{fig:250320_RT_PCCM}
\end{figure}

where $\alpha$ is a weight parameter and $f$ is the target value for $F_{AB}$. Minimizing this loss function optimizes the cloner towards a given target fidelity $f$. The target value for $f$ is chosen such that the eavesdropper can evade detection, while simultaneously maximizing the eavesdropping fidelity $F_{AE}$. For demonstration purposes, the PCCM attack (see Figure~\ref{fig:BB84_circ_meas}) is used and the attack angle $\theta$ is optimized using QCL. Figure~\ref{fig:250320_RT_PCCM} shows the training result on the trapped-ion hardware backend, using COBYLA~\cite{powell1994cobyla} as the classical optimizer. Additionally, we show the results of an ideal simulation of each iteration during the optimization. The target value was chosen to be $f=0.85$, close to the optimal attack. As for all data shown above, the correlations are reduced with respect to the ideal values due to circuit-level noise. Even though the final result of the loss functions deviates from the target due to imperfect operation of the QPU, the corresponding trained $\theta$ gives a result close to the target fidelity when evaluated without circuit level noise.

\subsubsection{BBM92}
\begin{figure*}[!htp]
    \centering
    \includegraphics[width=0.65\textwidth]{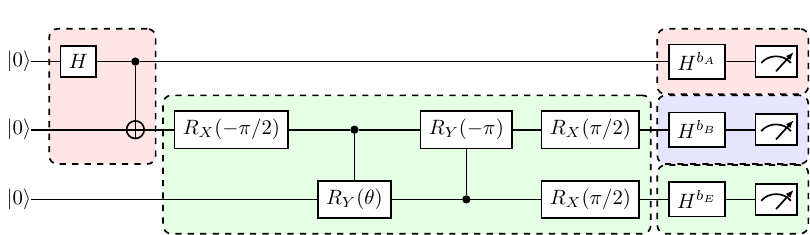}
    \caption{Circuit for the emulation of the BBM92 protocol between A (red) and B (blue). A PCCM attack by E (green) is performed. A Bell state is prepared and shared among A and B. E intercepts the qubit sent to B and clones the state to a blank qubit. Finally, all parties perform $H$ rotations depending on the basis choice and a projective measurement.}
	\label{fig:BBM92_circ_meas}
\end{figure*}

\begin{figure}[!htp]
    \centering
    \includegraphics[width=0.7\columnwidth]{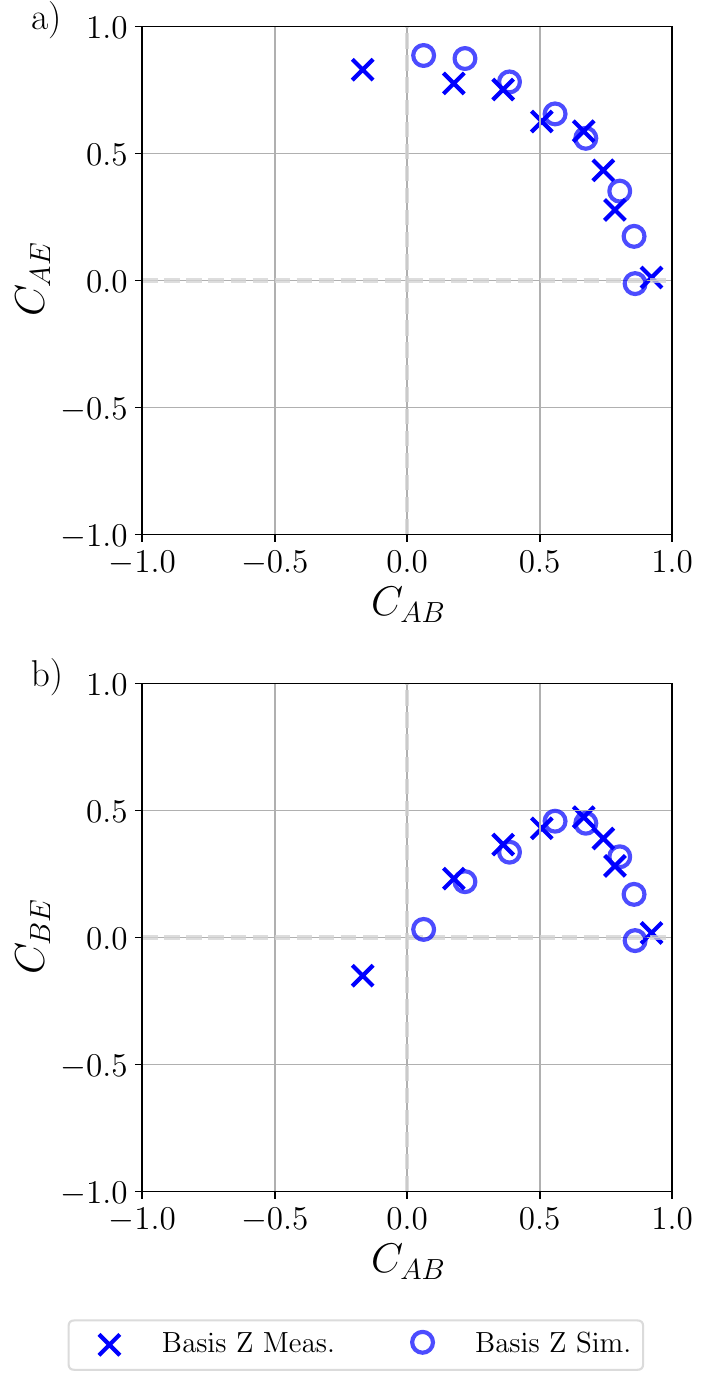}
    \caption{
    Correlations in case of a PCCM attack applied to the BBM92 protocol. The employed circuit is shown in Fig.~\ref{fig:BBM92_circ_meas}. \textbf{a)} Measured correlation estimates $C_{AE}$ vs. $C_{AB}$ for a symmetric, error free channel. We retrieve the behavior of a symmetric cloner, showing performance identical to the PCCM results displayed in Fig.~\ref{fig:BB84_PCCM_no_err} based on the BB84 protocol. \textbf{b)} Measured correlation estimates $C_{BE}$ vs. $C_{AB}$ for a symmetric, error free channel. A number of 1000 identical runs were used for bit 0 and Z basis configuration. Numerical simulations (circles) were performed with circuit-level depolarizing noise of strength $1\%$ on all gates (see Eqs.~\eqref{eq:depol} and \eqref{eq:depol_circ}).}
	\label{fig:BBM92_PCCM_no_err}
\end{figure}

In this section, we show the extension of the emulation of QKD protocols using the QI-Nutshell approach to the BBM92 protocol described in Sec.~\ref{sec:bbm92intro}. The emulation is based on three trapped-ion qubits. The emulation circuit is shown in Fig.~\ref{fig:BBM92_circ_meas}. A prepares an entangled Bell state and attempts to route one of the qubits to B. However, C intercepts the qubit and uses it as input for a PCCM before rerouting it to B. Finally, all three parties measure their qubit, where E delays their basis choice until the public information on the basis used by Alice is available. Here, the demonstration is restricted to the case without injected noise and one preparation and measurement basis only. The measured correlation estimates resulting from execution of the protocol are shown in Fig.~\ref{fig:BBM92_PCCM_no_err}. The joint correlations $C_{AB}$ and $C_{AE}$ show similar behavior as for the PCCM attack on the BB84 protocol shown in Fig.~\ref{fig:BB84_PCCM_no_err}. We also show $C_{BE}$ versus the estimated correlation $C_{AB}$, again for different attack angles. It can be observed that the correlation between B and E peaks at an intermediate, optimum attack angle, while the correlations vanish in the limit of a too strong attack, where the entanglement between A and B is swapped to A and E.

\subsection{Side-Channel Attacks}

The security of QKD protocols was extensively studied over the last decades~\cite{renner2006securityquantumkeydistribution, rusca2024quantumcryptographyoverviewquantum}. Currently, some QKD protocols are information-theoretically secure by means of proofs relying on restrictive assumptions~\cite{Renner_2005, wiesemann2024consolidatedaccessiblesecurityproof}. However, for a physical implementation of a QKD protocol to be validated as secure, it is not sufficient to prove the security of the protocol itself, but the vulnerability of specific implementations also has to be considered.

In Ref.~\cite{tupkary2025qkdsecurityproofsdecoystate} it is argued that there is currently no publication on the security of QKD that covers all security criteria that have to be met in a real application. The relevant literature is vast, scattered and only considers partial aspects of security with potential mismatches between assumptions and implementations. Additionally, the effect of side-channels on QKD protocols yet remains under-explored and it is generally unknown to which extent device imperfections may affect security guarantees. This is especially true if multiple side-channels are exploited simultaneously. Research is advancing and certain side-channels have been addressed, most recently in Refs.~\cite{sixto2025quantumkeydistributionimperfectly, trefilov2024intensitycorrelationsdecoystatebb84, nahar2025imperfectdetectorsadversarialtasks}, but still have to be thoroughly characterized beyond examples for specific hardware implementations. Ultimately, QKD is aiming at device-independent (DI) security. To the best of our knowledge, no real-life demonstration has been performed and security proofs are still under scrutiny. For a review on DI-QKD, the reader is referred to Ref.~\cite{ghoreishi2025futuresecurecommunicationsdevice}.
A review on the numerous side-channel attack strategies for QKD implementations exploiting hardware imperfections can be found in Ref.~\cite{BSI_side_channel_attack_study}. In this work, we exploit the ability to control the trapped-ion emulator at the level of hardware operations. This means that we leverage the ability to implement operations from the toolbox provided by atomic physics, in order to emulate building blocks of side-channel attacks. In the following, two such processes are discussed and characterized.

\subsubsection{Leakage of measurement results}
\label{sec:sideattacks:measurement_leakage}

A potential loophole for side-channel attacks is the ability of E to gain knowledge about the measurement outcome obtained by B. Such a process can be implemented in our setup by allowing Eve to control the processing zone after Bob's measurement. To extract information about B's measurement, E can perform a second state detection after B has finished his measurement. The amount of information E can extract can be controlled by varying the duration of Eve's detection pulse. This way, we can emulate realistic side-channels, which generally only admit an incomplete and imperfect extraction of information.

\begin{figure}[htp!]
    \centering
    \includegraphics[width=1.0\columnwidth]{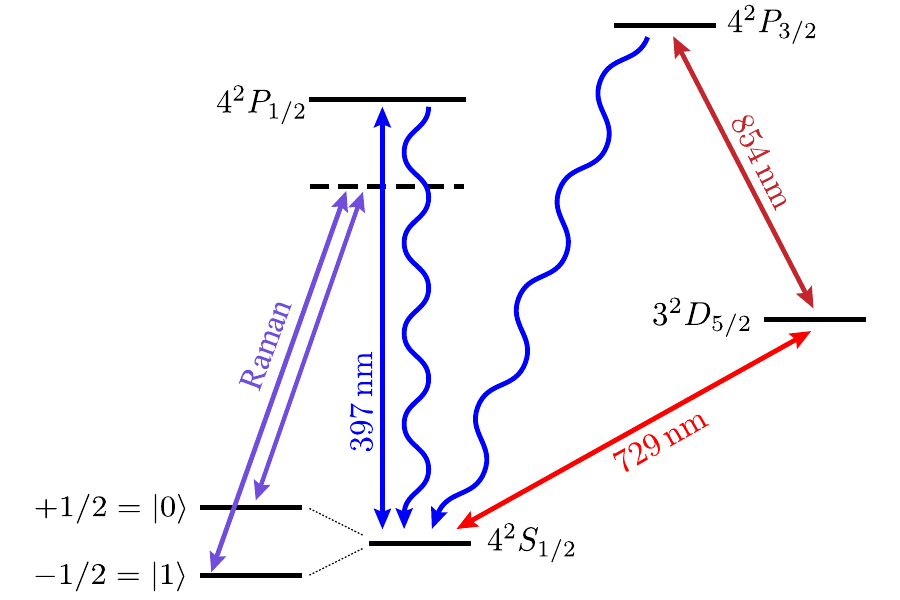}
    \caption{Energy level scheme of $^{40}$Ca$^+$. The qubit states are encoded in the two Zeeman sublevels of the $4^2S_{\nicefrac{1}{2}}$ ground state, Zeeman sublevels for the other state manifolds are not shown. Lasers at wavelengths of \SI{397}{\nano\m}, \SI{729}{\nano\m} and \SI{854}{\nano\m} are used to drive transitions between different electronic states. The qubit states are manipulated via a stimulated Raman transition with a detuning on the order of \SI{800}{\giga\Hz} from the $4^2S_{\nicefrac{1}{2}} \leftrightarrow 4^2P_{\nicefrac{1}{2}}$ transition.}
	\label{fig:level_scheme}
\end{figure}

In our experimental setup we use an auxiliary technique termed \emph{electron shelving} for projective measurements of the qubits in the Z basis~\cite{Poschinger2009}. The first step of the detection procedure is to transfer population in the state $\ket{0}$ encoded in the electronic state $S_{\nicefrac{1}{2}}, m_j=\nicefrac{1}{2}$ of $^{40}$Ca$^+$ to the state $D_{\nicefrac{5}{2}}$ (see energy level diagram in Fig.~\ref{fig:level_scheme}). Subsequently, the ion to be measured is illuminated with laser near \SI{397}{\nano\m}. Ions projected to $\ket{1}$ encoded in $S_{\nicefrac{1}{2}}, m_j=-\nicefrac{1}{2}$ scatter resonance fluorescence photons from the laser field, that are collected via a photomultiplier tube. In contrast, an ion projected to the state $\ket{0}$, which was transferred to $D_{\nicefrac{5}{2}}$, does not scatter photons.

The amount of information leaked about B's measurement results can be tuned via the duration of the side-channel measurement. If the exposure time for E's side-channel measurement is sufficiently long, near-perfect correlation with B's measurement will be obtained, irrespective of the qubit state and up to SPAM errors. For decreasing exposure time for E's measurement, an insufficient average number of photons will lead to an increasing rate of false-dark events, i.e. wrong assignments of measurement result $\ket{0}$ to qubit state $\ket{1}$. Therefore, the exposure time of the side-channel measurement can serve for tuning the correlation between the side-channel result of E and B's result.

\begin{figure}[!htp]
    \centering
    \includegraphics[width=\columnwidth]{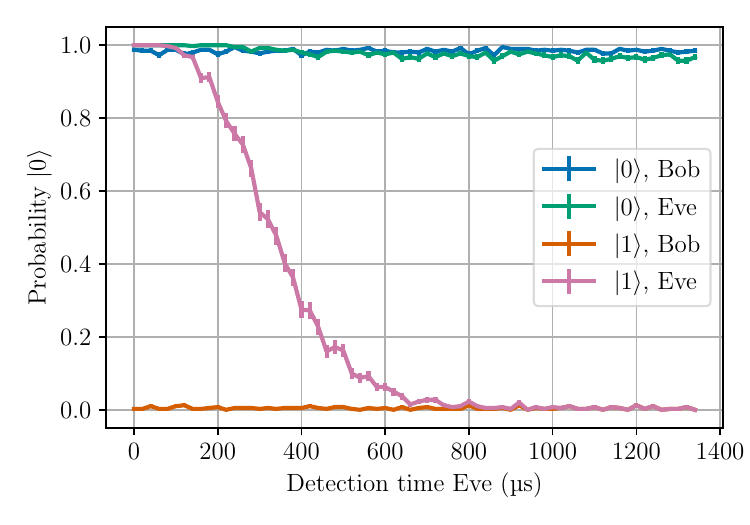}
    \caption{E uses a second state detection pulse after B's detection to gain information on B's measurement outcome. E's probabilities of detecting the ion as dark are shown in green and magenta for a qubit prepared in $\ket{0}$ and $\ket{1}$, respectively. The similarity to B's measurement outcomes, shown in purple and blue, increases with the duration of E's detection pulse. Each data point corresponds to 400 shots.}
	\label{fig:doubledetect}
\end{figure}

Figure~\ref{fig:doubledetect} shows Bob's and Eve's probabilities to detect an ion as 'dark' and consequently assign the result $\ket{0}$, versus the duration of Eve's detection. For B, a constant exposure time of \SI{1100}{\micro\second} was used. Data was taken for input states $\ket{0}$ and $\ket{1}$. It can be seen how E's results increasingly deviate from B's results for decreasing exposure times below \SI{100}{\micro\second}.

A side-channel attack using a similar information leakage channel is the \emph{breakdown flash} attack~\cite{BSI_side_channel_attack_study}. When exploiting the breakdown flash in a QKD setup, E detects photons that are emitted by B's detector back to the quantum channel when a photon is detected.

\subsubsection{Biasing measurement outcomes}

\begin{figure}[!htp]
    \centering
    \includegraphics[width=\columnwidth]{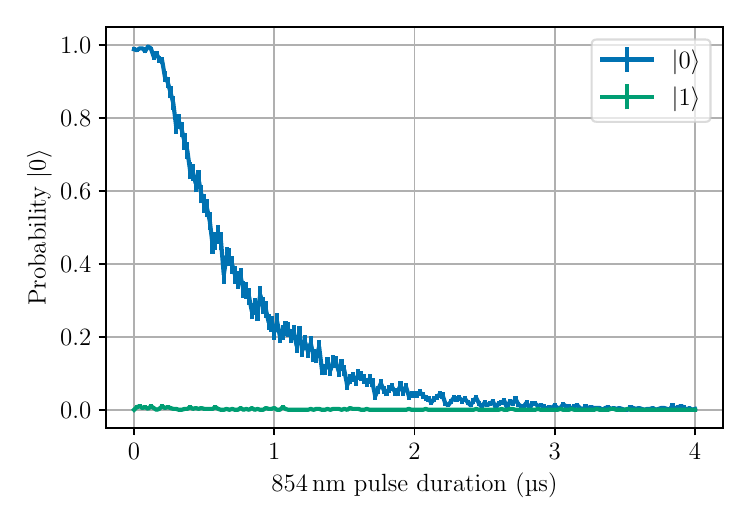}
    \caption{E uses a quench laser pulse at \SI{854}{\nano\m} to decrease B's probability to detect the ion as dark. The blue and green markers show Bob's probability to detect the ion as dark for the input states $\ket{0}$ and $\ket{1}$, respectively. For quench pulses with durations beyond \SI{3}{\micro\second}, most of the population in $D_{\nicefrac{5}{2}}$ is returned to the ground state and detected as bright. Each data point corresponds to 400 shots.
    }
	\label{fig:quench}
\end{figure}

Apart from gaining information about B's measurement, E could also actively affect B's measurement outcome. Within a QI-Nutshell emulation, by gaining control over the quantum processing zone before B's measurement, E can bias the outcomes that B obtains. If E illuminates the ion with a laser pulse at \SI{854}{\nano\meter}, referred to as a \emph{quench} pulse, after B shelved the population from the state $\ket{0}$ to the $D_{\nicefrac{5}{2}}$ manifold for qubit state discrimination, the shelved population is pumped back to $S_{\nicefrac{1}{2}}$ via excitation to $P_{\nicefrac{3}{2}}$~\cite{Poschinger2009}. As a result, the part of the population that was pumped from $D_{\nicefrac{5}{2}}$ to $S_{\nicefrac{1}{2}}$ cannot be distinguished from the leftover ground-state population in $\ket{1}$ after the shelving process, consequently the probability for B to detect the ion as bright increases. Figure~\ref{fig:quench} shows the dark-detection probability versus the duration of the quench laser pulse. For quench pulse durations beyond \SI{3}{\micro\second}, B's probability to detect the ion as dark is close to zero, irrespective of the input state.

\begin{figure}[!htp]
    \centering
    \includegraphics[width=\columnwidth]{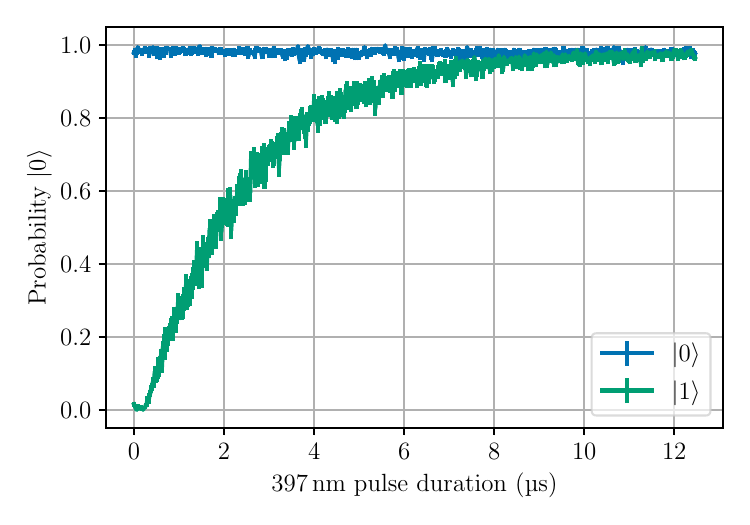}
    \caption{Eve applies a circularly polarized laser pulse near \SI{397}{\nano\m} before B's shelving operation to increase B's probability to detect the ion as dark. The plot shows B's probabilities to detect the ion as dark versus the duration of the pumping pulse, for input states $\ket{0}$ (blue) and $\ket{1}$ (green). Beyond pumping durations of \SI{10}{\micro\second}, the probability for Bob to successfully detect the ion in the prepared state $\ket{1}$ is close to zero. Each point corresponds to 400 shots.
    }
	\label{fig:sigma_pump}
\end{figure}

On the other hand, by illuminating the ion with circularly polarized light near \SI{397}{\nano\meter} before B applies the shelving operation, E can controllably reduce B's probability to measure an ion as bright. The laser pulse pumps population from $\ket{1}$ to $\ket{0}$~\cite{Poschinger2009}. Population in $\ket{0}$ is subsequently transferred to the $D_{\nicefrac{5}{2}}$ state via the shelving operation. Therefore, with increasing \SI{397}{\nano\m} pulse duration, B's probability to detect the ion as dark increases. As can be seen in Fig.~\ref{fig:sigma_pump}, beyond a circularly polarized \SI{397}{\nano\meter} pulse duration of around \SI{10}{\micro\second} B's dark detection probability is close to 100\%, regardless of the input state.

These emulated side-channel mechanisms for biasing of Bob's measurement outcomes resemble the mechanism used in \emph{detection efficiency mismatch attacks}~\cite{BSI_side_channel_attack_study}. In this attack scenario, E takes advantage of the fact that B's detectors may have different probabilities to detect an incoming photon, depending on a certain parameter, e.g. the photon arrival time~\cite{makarov2006effects}. E exploits this probability mismatch by manipulating this parameter. With this, Eve can make an impact on which of Bob's detectors clicks.


\section{Quantum Error Correction for Quantum Communication in the NISQ-Era}\label{sec:qec}

In a quantum network setting with noisy quantum communication channels, errors are introduced on individual qubits upon transmission~\cite{kimble2008quantum, van2014quantum, caleffi2018quantum}. A common assumption is that individual nodes in a quantum network are capable of performing QEC to remove such errors and restore the noise-free logical qubit states~\cite{slussarenko2022quantum, azuma2023quantumrepeaters, Shi_2024, hu2024quantum}. 
In this section, we aim to examine the practical use of single instances of small QEC codes in NISQ devices in a communication setup through numerical simulations.

We investigate how two small QEC codes can help to reduce error rates in quantum communication protocols and improve their reliability. We assume a generic situation where physical messenger qubits are first encoded into logical qubits in a QEC code. Upon reception, the encoded qubits are analyzed and afterwards decoded again. Our first example, the $[[4,2,2]]$ quantum error detection (QED) code~\cite{vaidman1996error}, is shown to reduce the QBER via post-selection. The $[[7,1,3]]$ Steane code~\cite{steane1996multiple} serves as a second example that illustrates how the distribution of measured syndromes after the quantum channel can allow one to detect deviations from an expected noise profile.
All numerical simulations were performed with \texttt{stim}~\cite{gidney2021stim} and \texttt{PECOS}~\cite{pecos}.

\subsection{Basic notions of quantum error correction}

We refer the reader to Refs.~\cite{nielsen2010quantum, devitt2013quantum} for a comprehensive introduction to QEC and only introduce some required terminology here. A quantum error correction code involves $n$ physical qubits that are prepared in a specific highly-entangled quantum state such that this state encodes $k \leq n$ computational degrees of freedom called \emph{logical qubits}.

Physical qubits must be considered noisy, as quantum states can generally be influenced by undesirable external disturbances. While noise could be minimized by isolating qubits from their environment, some sort of access to the qubits must inevitably be granted to an experimental control system in order to act with gates on the qubits and perform a computation. This trade-off places a lower limit on the noise-level being present in the system,
i.e., how reliably the physical qubit gates can be executed at most. By this mechanism, the maximally achievable circuit depth of a quantum computer working with physical qubits is constrained. 

Modern QEC routines typically work by measuring so-called \emph{stabilizer} operators, which do not alter the ideal, noise-free logical qubit state $\ket{\overline{\psi}}$ but instead diagnose the $n$-qubit quantum state for errors~\cite{gottesman1997stabilizer}. Any code state $\ket{\overline{\psi}}$ is a $+1$-eigenstate of any stabilizer operator and the $+1$-eigenspace of all stabilizers is called the code space. An erroneous code state $E \ket{\overline{\psi}}$ may be a $-1$-eigenstate to some stabilizer operator $g$ so that $g E \ket{\overline{\psi}} = - E g \ket{\overline{\psi}} = - E \ket{\overline{\psi}}$. In this case, we say that $E$ is a detectable error and it anticommutes with the stabilizer operator $g$. The classical bitstring that results from measuring the eigenvalues of a set of stabilizers is called the \emph{syndrome}. A syndrome bit is 0 if the corresponding stabilizer operator commutes with the error and it is 1 if the corresponding stabilizer operator anticommutes with the error.


For a QEC code with parameters $[[n,k,d]]$, the number of independent stabilizer operators is $n-k$ and $d$ is the \emph{distance} of the code, which reflects how many Pauli errors on a code state can be corrected at any point in time: QEC codes are capable of correcting $t = \lfloor (d-1) / 2\rfloor$ errors by inferring a feed-forward correction operation from any measured syndrome via a suitable \emph{decoder}. Alternatively, the distance-$d$ QEC code can be employed to detect $d-1$ errors at the price of introducing post-selection and thereby discarding some fraction of runs in a non-deterministic way. As a result, error rates will be suppressed from the physical error rate $p$ to a logical error rate proportional to $p^{t+1}$ or $p^{d}$ respectively in the low-$p$ limit. Remarkably, it has been shown that correcting Pauli errors is sufficient for a QEC code to correct noise in the form of arbitrary Kraus maps since the $n$-qubit Pauli operators form a complete basis~\cite{knill1997theory, nielsen2010quantum}.

In this work, we consider two types of potential noise models: In the \emph{code-capacity} noise model, one assumes that noise happens on the code level, i.e., on either of the $n$ physical qubits. 
This is the appropriate noise model for a noisy quantum communication channel that aims to transmit logical qubits, which are built from physical qubits. 
In the \emph{circuit-level} noise model, noise may happen on any $q$-qubit operation that acts on any subset of $q \leq n$ physical qubits as part of a quantum circuit, e.g., a single entangling gate. This noise model is generally considered more realistic in the context of fault-tolerant quantum computation. 


\subsection{Reduce QBER via the $[[4,2,2]]$ quantum error detection code}

\begin{figure}
    \centering
    \includegraphics[width=0.66\columnwidth]{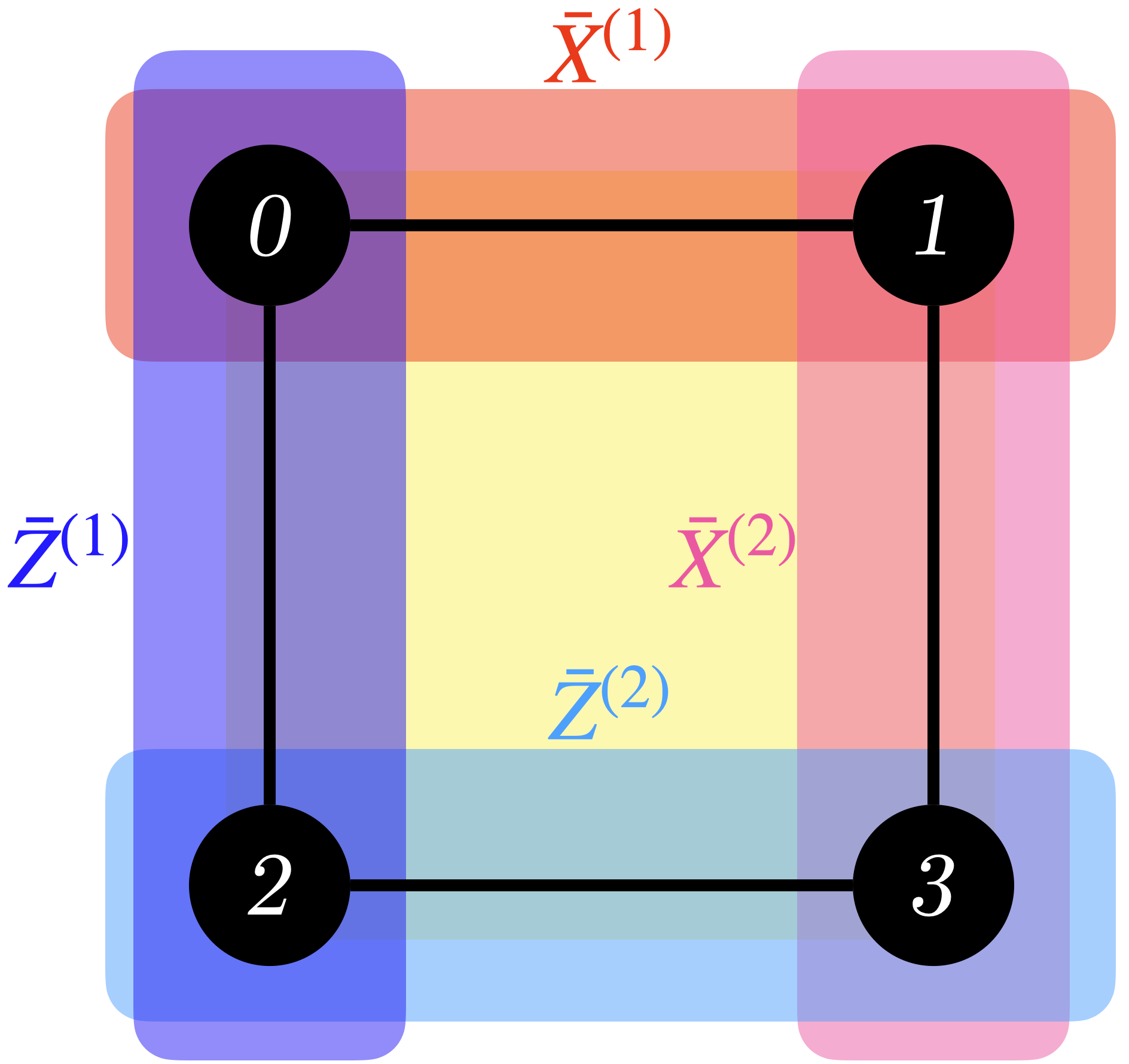}
    \caption{The $[[4,2,2]]$ code is an error detection code that can detect an arbitrary Pauli error $P_i \in \{X_i, Y_i, Z_i\}$ on any of its four physical qubits $i$ (black) by anticommutation with at least one of its stabilizers $g_X$ and $g_Z$ (yellow). Logical operators $\bar{X}^{(l)}$ (red) and $\bar{Z}^{(l)}$ (blue) for the same logical qubit $l = 1,2$ overlap on a single physical qubit.}
	\label{fig:422}
\end{figure}

The $[[4,2,2]]$ code~\cite{vaidman1996error} is a small QED code that has been implemented previously in several quantum computing hardware platforms~\cite{linke2017fault, takita2017experimental, erhard2021entangling, reichardt2024logicalcomputationdemonstratedneutral}. It can be depicted with a physical qubit arrangement such as in Fig.~\ref{fig:422}. The code space of the $[[4,2,2]]$ code is stabilized by the two operators 
\begin{align}
    g_X &= X^{\otimes 4}, ~~~~~~ g_Z = Z^{\otimes 4}
\end{align}
so that it defines $k = 2$ logical qubits. The logical operators can be chosen as
\begin{align}
    \bar{X}^{(1)} &= X_0X_1, ~~~~~~ \bar{Z}^{(1)} = Z_0Z_2,\notag \\
    \bar{X}^{(2)} &= X_1X_3, ~~~~~~ \bar{Z}^{(2)} = Z_2Z_3
\end{align}
so they have the same effect on the logical qubits as physical Pauli-$X$ and -$Z$ have on physical qubits. Also, the logical operators resemble the (anti)commutation relations of the physical Pauli operators since
\begin{align}
    \{\bar{X}^{(1)}, \bar{Z}^{(1)}\} &= 0, ~~~~~~ \{\bar{X}^{(2)}, \bar{Z}^{(2)}\} = 0 \notag \\
    [\bar{X}^{(1)}, \bar{Z}^{(2)}] &= 0, ~~~~~~ [\bar{X}^{(2)}, \bar{Z}^{(1)}] = 0
\end{align}
and all stabilizers commute with all logical operators. Any single physical Pauli operator anticommutes with at least one stabilizer and thus is indeed a detectable error. On the other hand, a weight-2 Pauli error\footnote{The weight of an error is the number of physical qubits it acts on non-trivially.}, such as $X_1X_2$, commutes with the stabilizers and is equivalent to a logical operator\footnote{Equivalent means that multiplication with stabilizer operators yields the same logical effect. Also, a product of logical operators is also a logical operator: Note that $\bar{X}^{(1)}\bar{X}^{(2)}$ takes the logical state $\ket{\overline{00}}$ to $\ket{\overline{11}}$. The error $X_1X_2$ can be multiplied by $X^{\otimes 4}$ to yield $X_0X_3 = \bar{X}^{(1)}\bar{X}^{(2)}$.}, here $\bar{X}^{(1)}\bar{X}^{(2)}$, which reflects the fact that the code has distance \mbox{$d=2$}.

Note that employing a distance-2 code does not allow one to distinguish exactly \textit{which} error has happened and, therefore, even a single-qubit error is not reliably correctable but can only be sorted out in post-selection. However, QKD can actually benefit from QED codes as QKD protocols rely on post-selection anyway; in contrast to fault-tolerant quantum computation, where mid-circuit measurements and feed-forward corrections appear as necessary ingredients for scale-up. 

\begin{figure}
    \centering
    \includegraphics[width=\columnwidth]{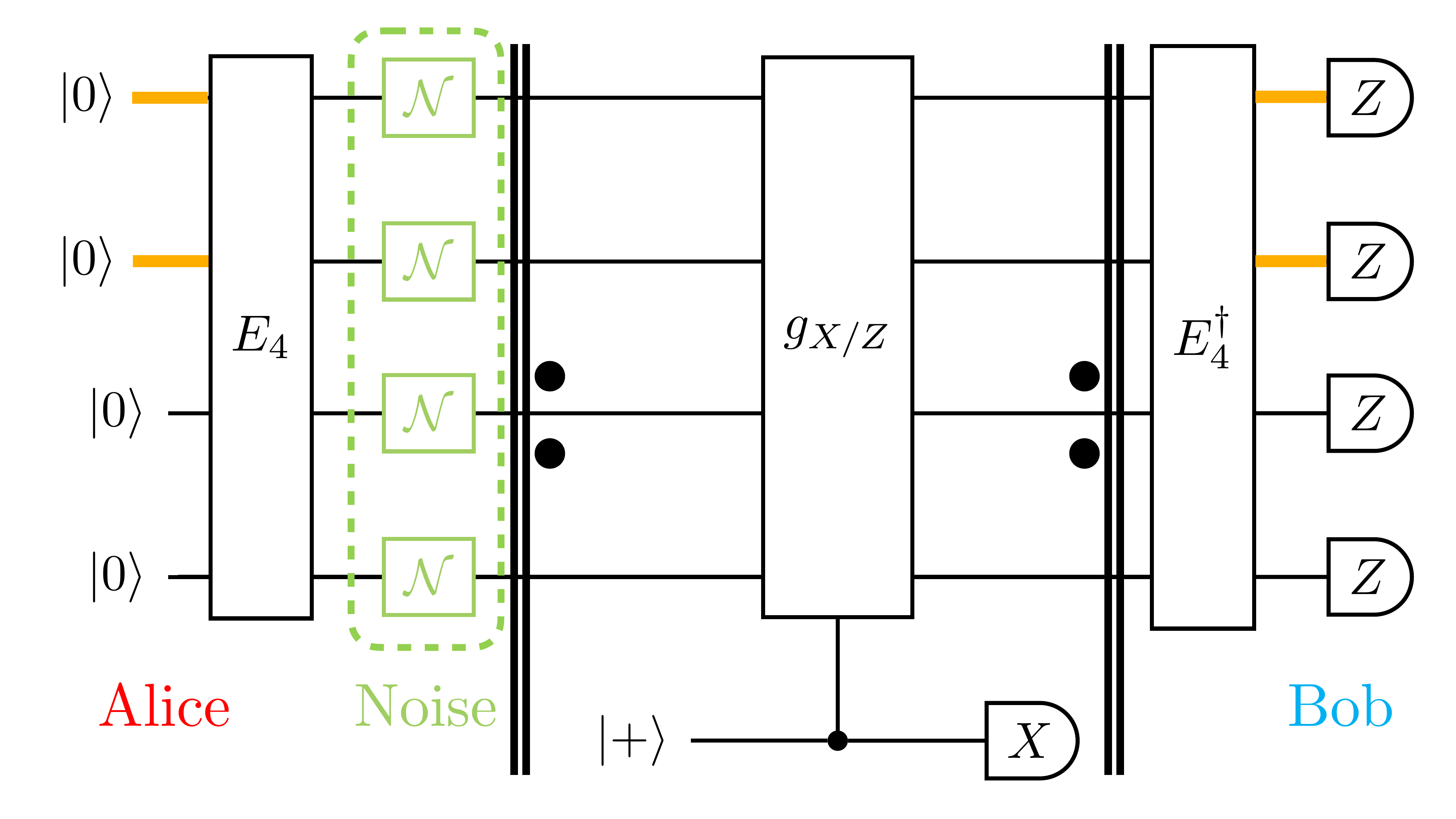}
    \caption{Circuit to perform error detection cycles with the $[[4,2,2]]$ code. A unitary encoding map $E_4$ is used to prepare the state $\ket{\overline{00}}$ starting from the state $\ket{0}^{\otimes 4}$. Input qubits are highlighted orange. Then, a channel noise map $\mathcal{N}$ acts four times independently on each physical qubit. An arbitrary but fixed number of repeated stabilizer measurements, decomposed into two-qubit gates, is performed next with the help of a single auxiliary qubit. In odd rounds we measure $g_X$ and in even rounds we measure $g_Z$. In the end, we apply the inverse encoding map $E_4^\dag$ and measure the physical qubits. The first two physical qubits carry the bit information. The last two physical qubits always end up in the state $\ket{0}$ in the absence of noise. Explicit circuits are given in Fig.~\ref{fig:422_enc}. Locations for circuit-level noise are \emph{not} shown explicitly.}
	\label{fig:422_circ}
\end{figure}

\begin{figure}
    \centering
    \includegraphics[width=\columnwidth]{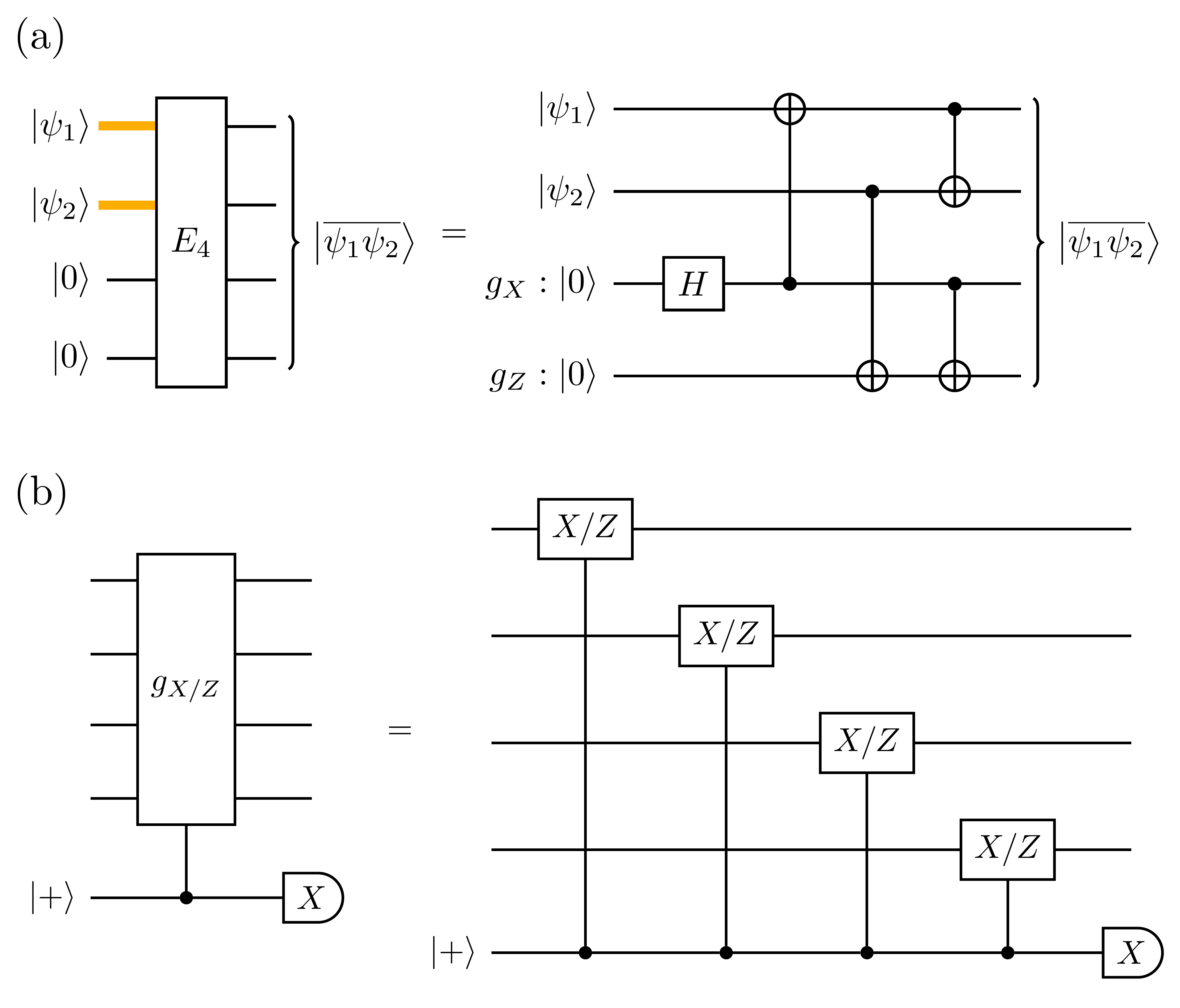}
    \caption{(a) Encoding circuit for the $[[4,2,2]]$ code. A single $Z$-operator on the third (fourth) qubit would propagate through the circuit to become the $g_X$ ($g_Z$) stabilizer. (b) Circuit to measure the stabilizer $g_X$ or $g_Z$ with the help of a single auxiliary qubit.}
	\label{fig:422_enc}
\end{figure}

\begin{figure*}
    \centering
    \includegraphics[width=\textwidth]{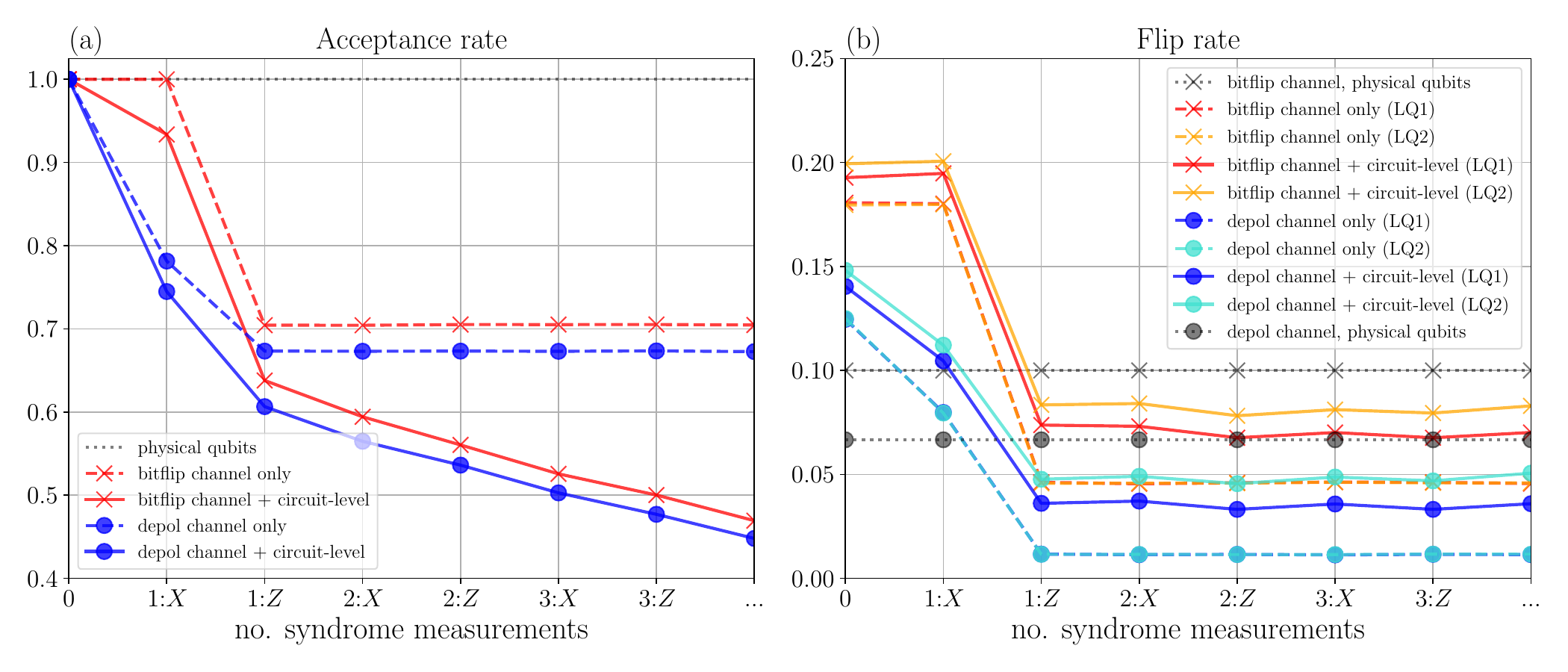}
    \caption{\textbf{Simulated rates for accepting a transmitted state and for retrieving a flipped qubit after repeated syndrome measurements.} We choose $\mathcal{N}$ to be either independent bitflip noise (Eq.~\eqref{eq:bitflip}, red, cross markers) or independent uniform depolarizing noise (Eq.~\eqref{eq:depol}, blue, circle markers) on any single qubit with respective strength $p = 0.1$ (see Fig.~\ref{fig:422_circ}). Dashed lines correspond to pure channel noise and solid lines correspond to channel noise and circuit-level depolarizing noise of strength $p_\mathrm{d} = 0.01$ on any single- or two-qubit operation. We take $10^6$ shots for each data point. \textbf{(a) Rate of trivial, i.e., $+1$ stabilizer measurement results.} In the absence of circuit-level noise, the first stabilizer measurement of $g_X$ fails to detect bitflip channel noise. $X$-flips can only be detected by the subsequent measurement of $g_Z$, contrary to depolarizing channel noise. Acceptance rates stay constant when repeating the stabilizer measurements. When adding circuit-level noise, the finite probability of $Z$-errors causes a decrease in acceptance rate for the first stabilizer measurement for both channels. 
    We observe a further decline in acceptance rate after the channel noise has been removed with one round of stabilizer measurements. 
    \textbf{(b) Ratio of flipped qubit outputs after post-selecting on the trivial syndrome.} With channel noise only, the flip rates of both 
    qubits (LQ1 and LQ2) coincide. We observe an overall increase of flip rates when adding circuit-level noise. After measuring both $g_X$ and $g_Z$ once, the flip rate stays approximately constant.}
	\label{fig:422_repmeas}
\end{figure*}

The circuit diagram in Fig.~\ref{fig:422_circ} reflects the procedure that we consider for using the $[[4,2,2]]$ code as part of the QKD pipeline. First, individual single qubits are encoded into a logical qubit state of the $[[4,2,2]]$ code via a unitary encoding map $E_4$. Here, we only consider encoding the physical qubit state $\ket{00}$ into the logical qubit state $\ket{\overline{00}} = \frac{1}{\sqrt{2}}\left(\ket{0000}+\ket{1111}\right)$. 
This is because syndrome measurements are entirely insensitive to which logical state they are acting on. So, the results would be the same in the case of, as for BB84, preparing $1, +$ or $-$ states. As a second step, the single-qubit noise map $\mathcal{N}$ is applied independently to each physical qubit and is supposed to model the noisy communication channel through which physical qubits are transferred from Alice to Bob. After transmission, a fixed 
number of stabilizer measurements are performed. We choose to measure $g_X$ and $g_Z$ in an alternating fashion with the help of a single physical auxiliary qubit that needs to be prepared in the $\ket{+}$ state and measured in the $X$-basis after performing a controlled-$g$ operation to map the stabilizer eigenvalue onto the measurement qubit~\cite{nielsen2010quantum, devitt2013quantum}. The corresponding physical circuits for these operations are shown in Fig.~\ref{fig:422_enc}. Lastly, the inverse encoding map is applied to retrieve the physical qubit states. Those are then measured in the $Z$-basis in order to determine the rate of bitflips\footnote{Note that there is additional syndrome information available from the extra physical qubits after $E_4^\dag$. We do not use these measurement results further in our analysis here.}. Only shots with the error-free, or trivial, syndrome are accepted and a shot is discarded whenever any syndrome measurement suggests that an error may have occurred.

\begin{figure}
    \centering
    \includegraphics[width=0.55\columnwidth]{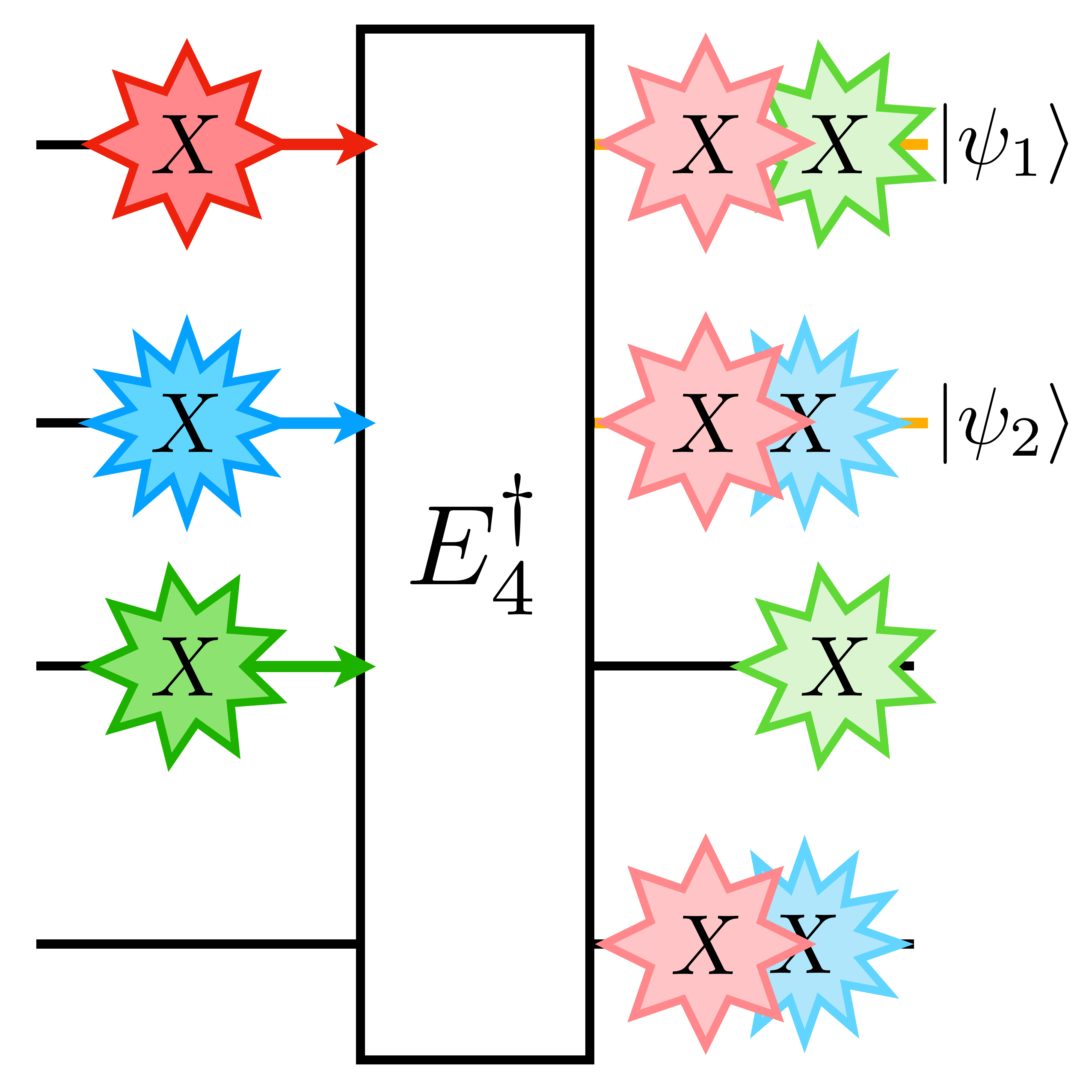}
    \caption{Errors that are present on the physical data qubits (stars with arrows) after stabilizer measurements (see Fig.~\ref{fig:422_circ}) cause flips of the qubit states $\ket{\psi_1}, \ket{\psi_2}$ by propagating through the inverse unitary encoder $E_4^\dag$ (see Fig.~\ref{fig:422_enc}). The errors $X_0$ (red) and $X_2$ (green) each flip $\ket{\psi_1}$, $X_0$ (red) and $X_1$ (blue) each flip $\ket{\psi_2}$.}
	\label{fig:422_dec}
\end{figure}

We take into account two prominent incoherent, Markovian noise maps in our analysis. The first is the bitflip channel
\begin{align}
    \mathcal{E}(\rho) = (1-p)\rho + p X \rho X, \label{eq:bitflip}
\end{align}
which applies the $X$-operator to a single-qubit state $\rho$ with a probability $p$ and leaves the state unchanged with complementary probability $1-p$. The second noise map applies uniform depolarizing noise
\begin{align}
    \mathcal{E}(\rho) = (1-p)\rho + \frac{p}{3} \left( X \rho X + Y \rho Y + Z \rho Z \right), \label{eq:depol}
\end{align}
with probability $p$ and leaves the state unchanged with probability $1-p$, i.e., the chance to apply an $X$-, $Y$- or $Z$-flip is $p/3$ each. Since the application of stabilizer measurements generally tends to decohere noise~\cite{beale2018quantum, iverson2020coherence}, we deem these channels sufficiently representative to model an exemplary noisy quantum communication channel. Additionally, we consider circuit-level depolarizing noise of the 
form in Eq.~\eqref{eq:depol}
for single-qubit operations and of the form
\begin{align}
    \mathcal{E}(\rho) = (1-p_\mathrm{d})\rho + \frac{p_\mathrm{d}}{15} \sum_{P \in \mathcal{P}_2 \backslash \{II\}} P \rho P \label{eq:depol_circ}
\end{align}
for two-qubit gates as well to account for faulty circuit operations that occur with noise strength $p_\mathrm{d}$. Here, $\mathcal{P}_2 \backslash \{II\}$ denotes the set of nontrivial two-qubit Pauli strings. It has been shown before that depolarizing noise can be sufficient to 
estimate the effect of circuit-level noise in real devices~\cite{postler2022demonstration, postler2024demonstration, pogorelov2025experimental}.

\begin{figure*}
    \centering
    \includegraphics[width=\textwidth]{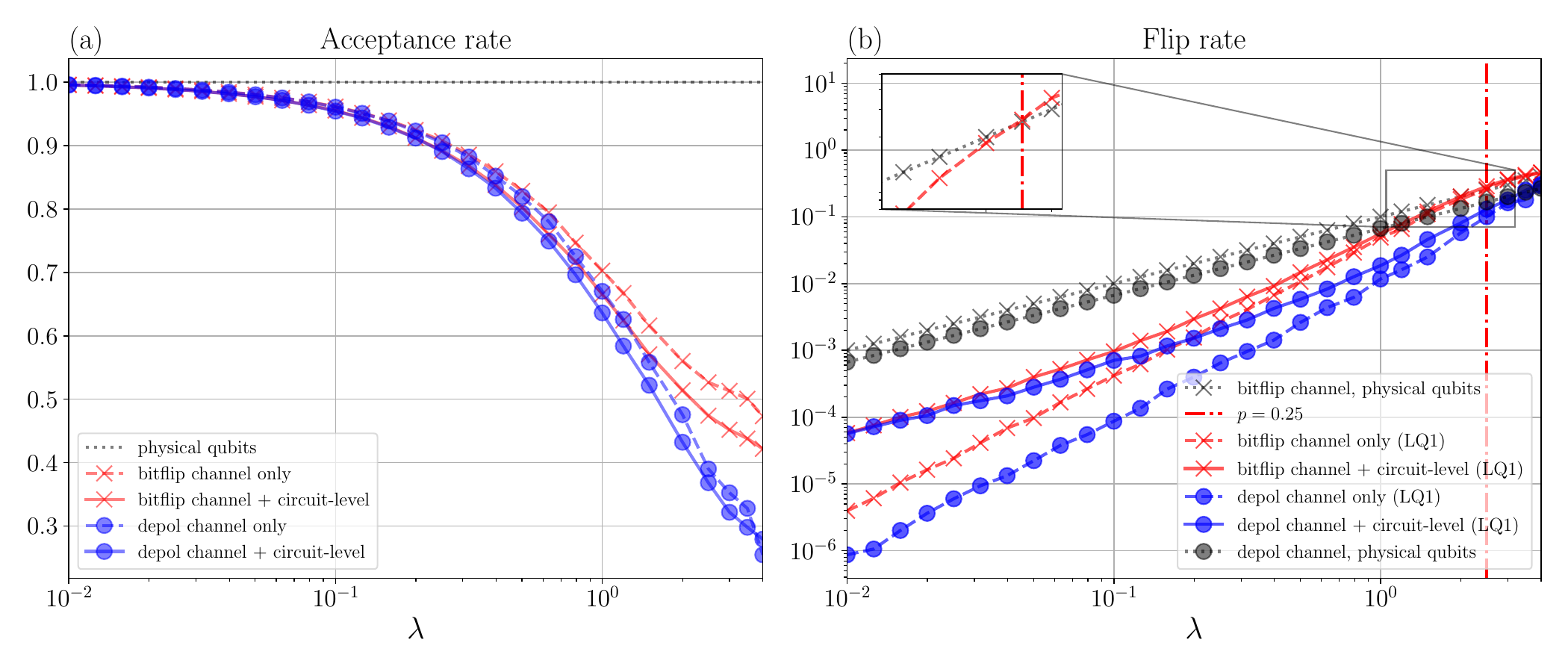}
    \caption{\textbf{Simulated rates for accepting a transmitted state and for retrieving a flipped qubit for varying noise strengths.} \textbf{(a)} \textbf{Acceptance rates decrease exponentially from unity for increasing noise strength.} We perform encoding, apply the specified noise channel $\mathcal{N}$, followed by the inverse encoding circuit and determine the syndrome based on the physical qubit measurement outcomes. This corresponds to one round of measuring each stabilizer (see Fig.~\ref{fig:422_enc}(a)). We choose a uniform noise-strength scaling parameter $\lambda$, i.e., the respective noise strengths are $\lambda p$ and $\lambda p_\mathrm{d}$. The point $\lambda = 1$ corresponds to $p = 0.1$ (and $p_\mathrm{d} = 0.01$). \textbf{(b)} \textbf{Scaling of qubit flip rates.} In the case of sole channel noise, we observe the scaling behavior $p_L = \mathcal{O}(p^2)$ but with circuit-level noise we observe $p_L = \mathcal{O}(p^1)$ in the low-$p$ limit ($\lambda \lesssim 10^{-1}$). The vertical line marks $p = 0.25$ below which we observe an advantage of logical qubits over physical qubits for the bitflip channel (inset). The physical qubit flip rate for comparison is $p(1-p) + p^2$ for the bitflip channel and $2p/3 \cdot (1-2p/3) + (2p/3)^2$ for the depolarizing channel. For the left-most data point we take $10^7$ shots for the depolarizing channel.}
	\label{fig:422_scaling}
\end{figure*}

It can generally be expected that only the first round of stabilizer measurements will have an effect in case of pure channel noise. Any detectable errors that occur after $E_4$, will be detected by either the first measurement of $g_X$ or $g_Z$. If an error is not detected here, it will also not be detected by any further measurement round. This can be observed in Fig.~\ref{fig:422_repmeas}. For the bitflip channel, the first round, where $g_X$ is measured, does not decrease the acceptance rate at all, since $g_X$ cannot detect $X$-errors (see Fig.~\ref{fig:422_enc}(b)). Only the subsequent measurement of $g_Z$ decreases the acceptance rate and the rate then stays constant, as expected. On the contrary, the depolarizing noise channel applies all three types of Pauli errors to the physical qubits. Therefore, here already the first measurement of $g_X$ detects $Y$- and $Z$-errors so that the acceptance rate already decreases in the first round. Adding a small but realistic amount of circuit-level depolarizing noise of strength $p_\mathrm{d} = 0.01$ (see Sec.~\ref{sec:experiments}) to all circuit operations on top as well, further decreases the acceptance rate due to additional errors in the encoding circuit and the stabilizer measurements. It also leads to errors being constantly detected in subsequent stabilizer measurements and a consistent decline of acceptance rates for both the bitflip and the depolarizing noise channel. The characteristic shape of the declining curve may allow us to make an educated guess about the nature of the noise channel that might not be known analytically in a realistic experimental setup.

For the bitflip channel, the noise floor in Fig.~\ref{fig:422_repmeas}(a) lies at a value of $1-(4 \cdot 0.1 \cdot (1-0.1)^3+4 \cdot 0.1^3 \cdot (1-0.1)) \approx 0.705$. This is the total probability that one or three $X$-errors occur. There are 4 distinct error configurations that each occur with the probability that 1 error happens multiplied with the probability that 3 qubits are error-free. Also, there are 4 distinct configurations where 3 errors happen and 1 qubit remains error-free, which are detectable and thus also contribute to the acceptance rate. There are no weight-2 and no weight-4 detectable errors since these commute with all stabilizers. The $[[4,2,2]]$ code, in this case, suppresses error rates as $p_L = 4p^2 + \mathcal{O}(p^3)$ so we achieve a lower error rate as long as $p_L < p \Rightarrow p \lesssim 1/4$.

On the right-hand-side of Fig.~\ref{fig:422_repmeas}, we scrutinize the effect of stabilizer measurements and post-selection on the resulting rate of flips on the transmitted qubit states, i.e., we record how many times we measured $\ket{1}$ instead of $\ket{0}$ for both qubits input to the circuit in Fig.~\ref{fig:422_circ}. For pure channel noise, the curves for the two qubits in Fig.~\ref{fig:422_repmeas}(b) lie on top of each other. As for the acceptance rate, the flip rate drops from its initial value after the first stabilizer measurement for the depolarizing channel and after the second stabilizer measurement for the bitflip channel.

Let us quantify the effect of the bitflip channel more explicitly: The initial qubit flip rate without stabilizer measurements is $2 \cdot 0.1 \cdot (1-0.1)^3+4 \cdot 0.1^2 \cdot (1-0.1)^2 + 2 \cdot 0.1^3 \cdot (1-0.1) = 0.18$ because we have two weight-1 errors that propagate to logical flips (see Fig.~\ref{fig:422_dec}). 
Additionally, each logical qubit can be flipped by four of the six undetectable weight-2 errors. Note that $X_0X_3 = \bar{X}^{(1)}\bar{X}^{(2)}$ and $X_1X_2 = g_X \bar{X}^{(1)}\bar{X}^{(2)} $ flip both logical qubits, $X_0X_1 = \bar{X}^{(1)}$ and $X_2X_3 = g_X \bar{X}^{(1)}$ only flip the first logical qubit and $X_0X_2 = g_X \bar{X}^{(2)}$ and $X_1X_3 = \bar{X}^{(2)}$ only flip the second logical qubit. Also, two weight-3 errors contribute that are stabilizer-equivalent to the aforementioned weight-1 errors. After having measured $g_Z$ for the first time, only the undetectable weight-2 errors remain on the logical qubit state and therefore the flip rate goes down to $4 \cdot 0.1^2 \cdot (1-0.1)^2 / 0.705 \approx 0.046$. Note that there might also be an undetectable weight-4 error, but since this is a stabilizer, no logical flip will occur. Such analysis by counting errors and quantifying their impact can be done analogously for the depolarizing noise channel.

The scaling behavior of acceptance rates and flip rates are of interest in the context of QEC. We introduce a parameter $\lambda$ to uniformly scale the respective noise strengths as $p \rightarrow \lambda p$ and $p_\mathrm{d} \rightarrow \lambda p_\mathrm{d}$. Our reference point $\lambda = 1$ corresponds to $p = 0.1$ and $p_\mathrm{d} = 0.01$. Figure~\ref{fig:422_scaling}(a) shows that all shots are accepted in the absence of noise, i.e., in the limit $\lambda \rightarrow 0$. Upon increasing noise strengths, the fraction of rejected shots grows exponentially. We observe the characteristic scaling $\mathcal{O}(p^2)$ in Fig.~\ref{fig:422_scaling}(b) as $p \rightarrow 0$ because the $[[4,2,2]]$ code can detect any of the 4 (12) single-qubit Pauli errors, which occur in $\mathcal{O}(p)$ for our bitflip (depolarizing) channel noise model. An advantage over physical qubits can be expected since their flip rates scale linearly as $p \rightarrow 0$. The probability that a given single physical qubit flips under the bitflip channel is determined as the probability that \emph{this} one qubit flips and the other does not or that both qubits flip at the same time $p(1-p) + p^2 = \mathcal{O}(p)$. The flip rate for physical qubits subjected to the depolarizing channel is analogously given as $2p/3 \cdot (1-2p/3) + (2p/3)^2 = \mathcal{O}(p)$.

In conclusion, we stress that using QEC codes in QKD may serve two purposes: We can infer noise characteristics from syndrome measurements and reduce the flip rate of a QKD transmission line via post-selection effectively, as we showcased with the $[[4,2,2]]$ quantum error detecting code as an example. The amount of noise suppression can be made quantitative through the known structure of the code.


\subsection{Monitoring noise via the $[[7,1,3]]$ Steane code}

\begin{figure}
    \centering
    \includegraphics[width=0.75\columnwidth]{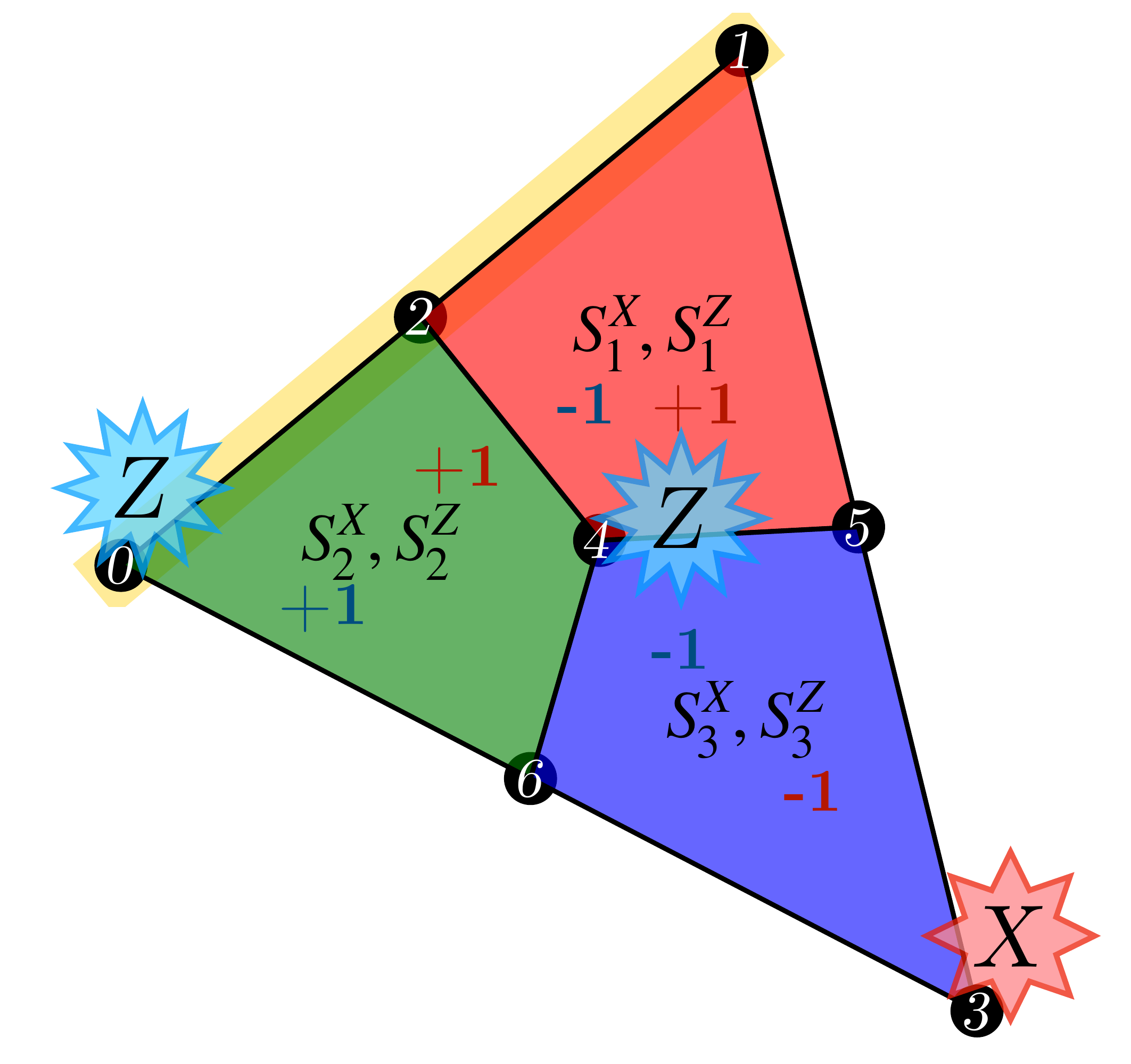}
    \caption{The $[[7,1,3]]$ Steane code is the smallest representative of the family of 2D topological color codes. It can correct one arbitrary Pauli error. Logical operators have minimal weight 3. Stabilizer generators have support on the weight-4 plaquettes and are symmetric under exchange of $X$ and $Z$. The error $X_3$ (red star) only anticommutes (-1) with $S_3^Z$ and therefore causes the $Z$-syndrome 001 (also see left-most bin in Fig.~\ref{fig:steane_res}). The error $Z_0Z_4$ (blue stars) is detectable because it flips $S_1^X$ and $S_3^X$ but commutes (+1) with $S_2^X$. It cannot be corrected if the $X$-syndrome 101 is already assigned to $Z_5$ (see Tab.~\ref{tab:lut_steane}). Weight-3 errors that correspond to logical operators (yellow) are undetectable.}
	\label{fig:steane}
\end{figure}

In this section, we follow up on the notion of inferring noise characteristics from syndrome measurements. We suggest using this syndrome information to monitor the noise profile of a quantum communication channel. For this purpose we employ a QEC code that provides longer syndromes with supposedly more (useful) information.

The $[[7,1,3]]$ Steane code~\cite{steane1996multiple} is the smallest representative of the code family of two-dimensional topological color codes~\cite{bombin2006distillation, bombin2006topological}. Since it has distance $d=3$, it is a QEC code that serves to correct $t = 1$ arbitrary Pauli error on any of its $n=7$ physical qubits that are encoded into $k = 1$ logical qubit. When using the Steane code for mere QED, any one or two errors can be detected by measuring the stabilizers
\begin{align}
    S_1^{X} &= X_1X_2X_4X_5, ~~~~~~ S_1^{Z} = Z_1Z_2Z_4Z_5 \notag \\
    S_2^{X} &= X_0X_2X_4X_6, ~~~~~~ S_2^{Z} = Z_0Z_2Z_4Z_6 \notag \\
    S_3^{X} &= X_3X_4X_5X_6, ~~~~~~ S_3^{Z} = Z_3Z_4Z_5Z_6.
\end{align}
Assigning a syndrome to a unique correction operation is only possible for the weight-1 errors and one may choose the following \emph{look up table} decoding strategy:
\begin{table}[h]
\centering
\begin{tabular}{|c|c||c|}
\hline
Syndrome & Correction & High-weight error\\ \hline
000 & $I$ & $P_0P_1P_2$ \\ \hline
001 & $P_3$ & $P_0P_6$, $P_4P_5P_6$ \\ \hline
010 & $P_0$ & $P_1P_2$, $P_2P_4P_6$\\ \hline
011 & $P_6$ & $P_1P_4$, $P_0P_2P_4$\\ \hline
100 & $P_1$ & $P_3P_5$, $P_2P_4P_5$\\ \hline
101 & $P_5$ & $P_0P_4$, $P_1P_2P_4$\\ \hline
110 & $P_2$ & $P_3P_4$, $P_1P_4P_5$\\ \hline
111 & $P_4$ & $P_2P_3$, $P_3P_5P_6$\\ \hline
\end{tabular}
\caption{Look up table for the Steane code. A unique single-qubit correction $P \in \{X, Z$\} can be determined by the independent but symmetric $Z$- or $X$-syndromes respectively (two left-most columns). This is a special property that follows from the Steane code being a self-dual CSS code. Higher weight errors may cause the same syndrome as a single-qubit error and are therefore uncorrectable (examples in right column). There exist undetectable weight-3 errors (see Eq.~\eqref{eq:logs}).}
\label{tab:lut_steane}
\end{table}

The logical operators of the distance $d=3$ code
\begin{align}
    \overline{X} &= X_0X_1X_2, ~~~~~~ \overline{Z} = Z_0Z_3Z_6 \label{eq:logs}
\end{align}
have minimal-weight three and can be viewed as acting along the boundaries of the code patch shown in Fig.~\ref{fig:steane}. Since all weight-1 and weight-2 errors are detectable by stabilizer measurements, the Steane code suppresses error rates asymptotically as $p_L = \mathcal{O}(p^3)$ with post-selection (or as $p_L = \mathcal{O}(p^2)$ with deterministic feed-forward corrections). Experimental implementations of the Steane code have been successfully demonstrated across a wide range of quantum computing hardware platforms~\cite{nigg2014quantum, ryan2021realization, hilder2022fault, bluvstein2022aquantum, 
bluvstein2024logical, lacroix2024scalinglogiccolorcode}.

\begin{figure}
    \centering
    \includegraphics[width=\columnwidth]{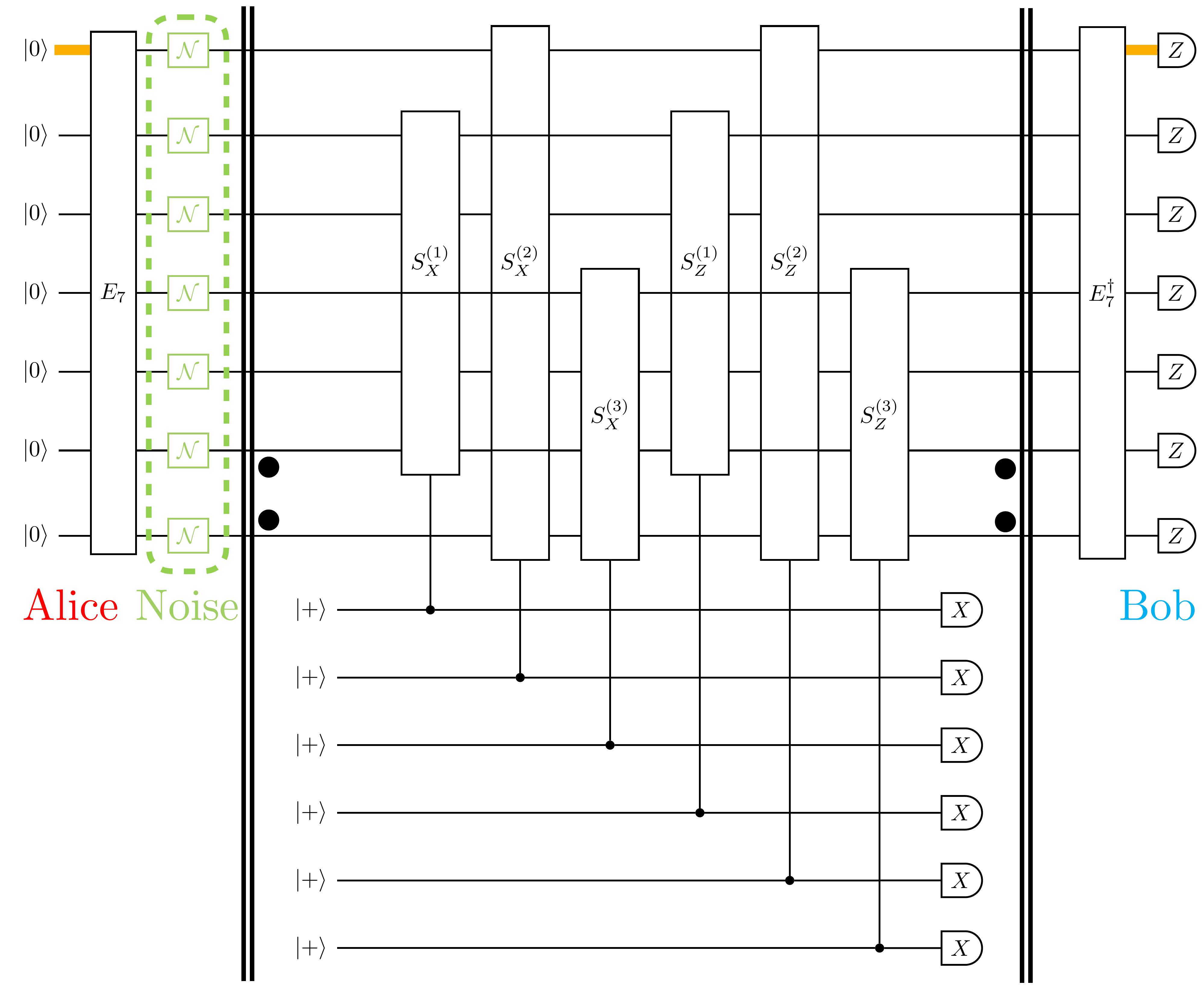}
    \caption{Circuit to perform rounds of syndrome measurements with the $[[7,1,3]]$ code. A unitary encoding map $E_7$ is used to prepare the logical state $\ket{\bar{0}}$ starting from the state $\ket{0}^{\otimes 7}$. Then, a channel noise map $\mathcal{N}$ acts 
    independently on each of the seven physical qubits. A fixed 
    number of repeated syndrome measurements, decomposed into two-qubit gates analogous to Fig.~\ref{fig:422_enc}(b), is performed next with the help of six auxiliary qubits (note that a single auxiliary qubit with intermediate re-initialization would suffice). Each auxiliary qubit is measured to obtain the eigenvalue of a single stabilizer generator. In the end, we apply the inverse encoding map $E_7^\dag$ and measure the physical qubits. The first physical qubit carries the bit information. The last six physical qubits always end up in the state $\ket{0}$ in the absence of noise. Locations for circuit-level noise are \emph{not} shown explicitly.}
	\label{fig:steane_circ}
\end{figure}

Let us now focus on employing the six-bit syndromes of the Steane code in order to monitor our noisy quantum communication channel. For this purpose we assemble a circuit for encoding, application of the channel noise map, repeated stabilizer measurements and decoding as illustrated in Fig.~\ref{fig:steane_circ}. Six physical auxiliary qubits are used in each round of measurements of the six-bit combined $X/Z$ syndrome. We showcase two variants of the depolarizing noise channel in Eq.~\eqref{eq:depol} that can be identified from the syndrome distribution. The first variant strongly amplifies the noise strength $p$ on a single physical qubit by a factor of 10. The second variant introduces a bias such that instead of applying $X$-, $Y$- and $Z$-flips with equal probability $p/3$, we apply $X$-errors with probability $p_X = 0.1$ and $Y$- and $Z$-errors with much smaller probabilities $p_Y = p_Z = 0.01$.

\begin{figure*}
    \centering
    \includegraphics[width=\textwidth]{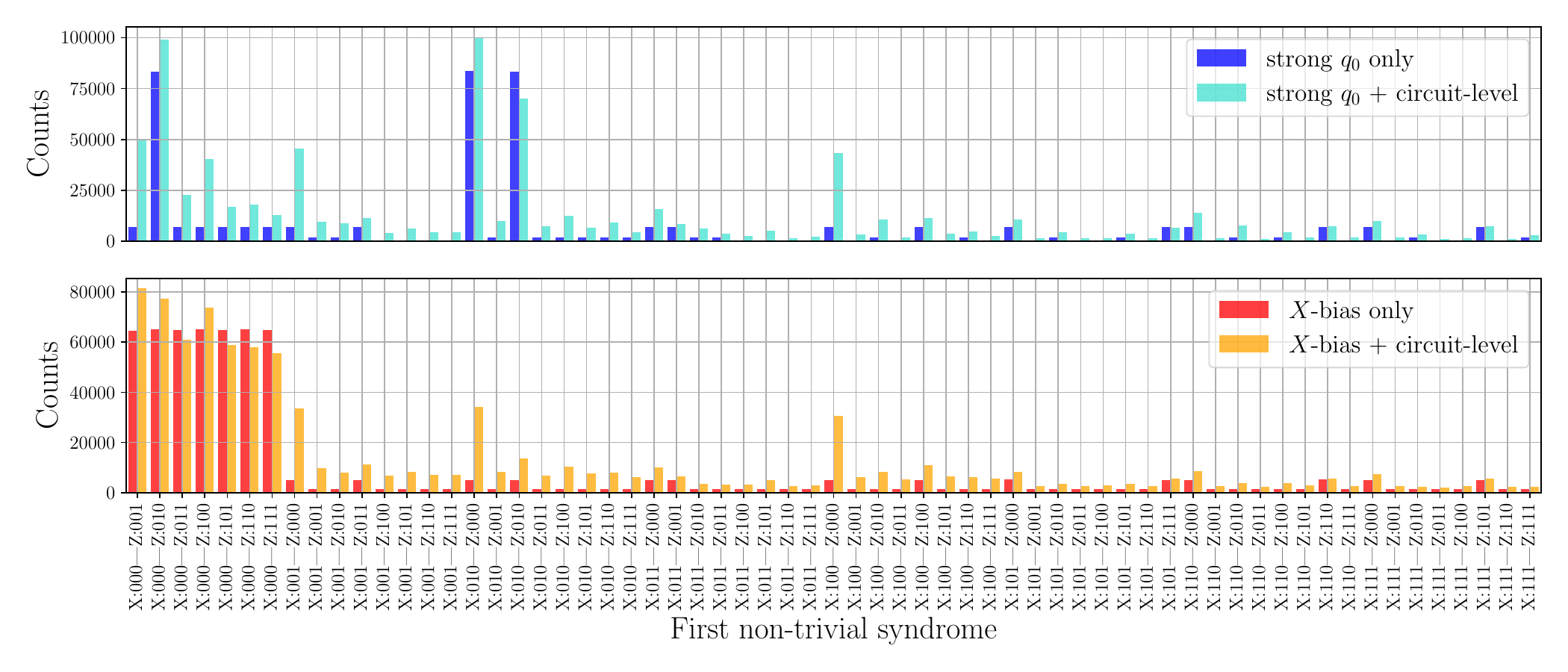}
    \caption{We perform up to three rounds of syndrome measurements in simulation with $10^6$ shots, as sketched in Fig.~\ref{fig:steane_circ} and record the first non-trivial syndrome that leads to the result being discarded in post-selection. \textbf{Upper}: We choose $\mathcal{N}$ to be a uniform depolarizing channel with strength $p = 0.03$ on qubits 1 to 6 and strength $10p = 0.3$ on qubit 0. The three syndromes $010\,000,010\,010$ and $000\,010$ that correspond to the errors $X_0$, $Y_0$ and $Z_0$ are recorded most often (blue), also in the presence of circuit-level noise (turquoise). \textbf{Lower}: We choose $\mathcal{N}$ to be a non-uniform Pauli channel with $p_X = 0.1$ and $p_Y = p_Z = 0.01$. The seven pure $Z$-syndromes are recorded most often (red), also in the presence of circuit-level noise (orange). 
    }
	\label{fig:steane_res}
\end{figure*}

The syndrome distributions for both variants, each without and with a small amount of additional circuit-level depolarizing noise ($p_\mathrm{d} = 0.01$, cf.~Figs.~\ref{fig:BB84_PCCM_no_err} and~\ref{fig:BB84_all_x}), in up to three rounds of measuring the full syndrome are shown in Fig.~\ref{fig:steane_res}. When a non-trivial syndrome bitstring is observed, we stop the run and record the syndrome. When a trivial syndrome is observed, we follow up with the next round of syndrome measurements.

For 
noise preferentially acting on qubit 0, one can clearly observe in Fig.~\ref{fig:steane_res} (upper panel) that syndromes that correspond to a single error on this qubit (see Tab.~\ref{tab:lut_steane}), are recorded for a majority of the shots. This peak structure in the syndrome distribution can be interpreted as a signature of the noise channel that allows one to directly infer that qubit 0 is noisier than the other qubits. Strikingly, this overall structure remains widely intact when we also add circuit-level depolarizing noise to the circuit. The three largest peaks indicating toward qubit 0 remain the most prominent peaks of the distribution.

For the other scenario of a bias towards Pauli $X$-errors on all qubits, we observe a different signature. Figure~\ref{fig:steane_res} shows that in this case (lower panel) the majority of the recorded syndromes are the seven non-trivial pure $Z$-syndromes while the $X$-part of the syndrome is trivial (000), as one would expect for pure $X$-errors on the seven data qubits (cf.~Tab.~\ref{tab:lut_steane}). Only a small fraction of runs yields other syndromes due to $p_Y$ and $p_Z$ also being non-zero. Again, adding a small but realistic amount of circuit-level noise on top ($p_\mathrm{d} = 0.01$) only contributes to the noise floor but retains the overall pattern.

To model a realistic QKD scenario, the channel noise maps we used so far may be too simplistic. For this reason we now introduce a more intricate noise model and show that the essential features of our setup outlined up to this point largely remain present.
For instance, the Kraus operators of the amplitude damping channel read
\begin{align}
E_1 &= \begin{pmatrix}
1 & 0 \\
0 & \sqrt{1 - \gamma}
\end{pmatrix}, ~~~~~~ E_2 = \begin{pmatrix}
0 & \sqrt{\gamma} \\
0 & 0
\end{pmatrix}
\end{align}
with a parameter $\gamma$. 
It is known that, upon Pauli twirling, the amplitude damping channel transforms into a non-uniform depolarizing channel
\begin{align}
    \mathcal{E}(\rho) &= \frac{2 + 2\sqrt{1 - \gamma} - \gamma}{4} \rho \notag \\ 
    &+ \frac{\gamma}{4} X \rho X 
    + \frac{\gamma}{4} Y \rho Y 
    + \frac{2 - 2\sqrt{1 - \gamma} - \gamma}{4} Z \rho Z, 
\end{align}
which we use for our simulation~\cite{katabarwa2017dynamicalinterpretationpaulitwirling}. Additionally, the 
phase of individual qubits 
may be altered by the transmission line. We explicitly take the pure dephasing channel
\begin{align}
    \mathcal{E}(\rho) = (1-p_\mathrm{pd}) \rho + p_\mathrm{pd} Z \rho Z
\end{align}
into account for our simulations with a parameter $p_\mathrm{pd}$. 

While the physical loss of individual qubits is difficult to correct, we only seek to detect it~\cite{vala2005quantum, devitt2013quantum}. 
We additionally model loss according to \texttt{stim}'s \texttt{HERALDED\_ERASE} functionality.

\begin{figure*}
    \centering
    \includegraphics[width=\textwidth]{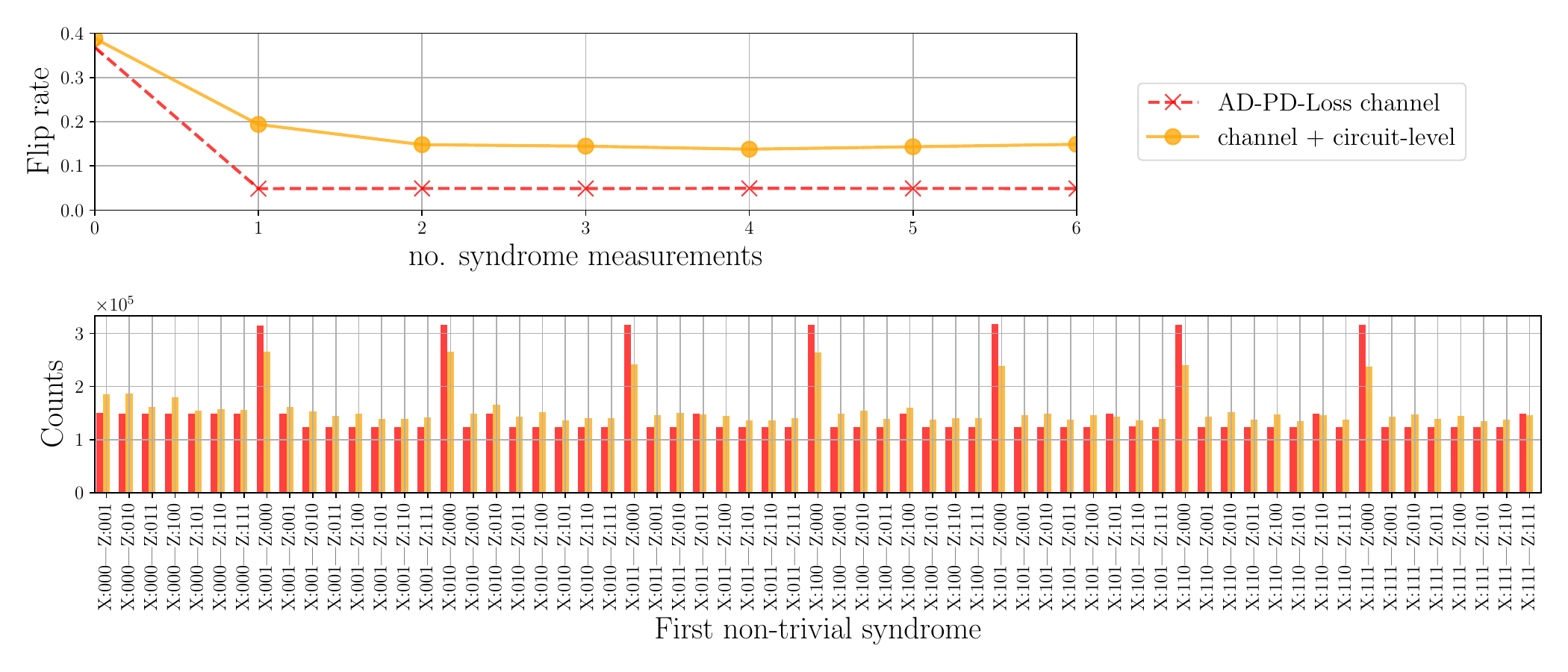}
    \caption{We perform up to six rounds of syndrome measurements in simulation with $10^7$ shots, similar to Fig.~\ref{fig:steane_circ}, and record the first non-trivial syndrome that leads to the result being discarded in post-selection. \textbf{Upper:} We choose $\mathcal{N}$ to be the subsequent application of an amplitude damping (AD) channel, a pure dephasing (PD) channel and \texttt{stim}'s \texttt{HERALDED\_ERASE} channel. For pure channel noise (red dashed line, cross markers), the flip rate drops to a plateau after one round of syndrome measurements. \textbf{Lower:} The syndrome distribution for pure channel noise (red) acts as a baseline of the expected measurement structure. Adding circuit-level noise alters the syndrome distribution (orange) but keeps the qualitative features intact.}
	\label{fig:real_res}
\end{figure*}

With the exact same protocol as before but this refined channel noise map where these three processes are applied sequentially, we again record the syndrome distribution with up to 
six rounds of syndrome measurements. The channel noise parameters are now set to $\gamma = p_\mathrm{pd} = p_\mathrm{l} = 0.2$. The characteristic peak structure of the pure channel noise is displayed in Fig.~\ref{fig:real_res}. 
Due to the strong prevalence of $Z$-errors and equal but lower probability for $X$- and $Y$-errors, we observe peaks at the pure $X$-syndromes and a relatively uniform noise floor.
For the syndrome distribution, 
the largest peaks from the situation of pure channel noise remain intact when 
adding circuit-level noise of strength $p_\mathrm{d} = 0.025$, which may be present in 
NISQ devices. So even in this more elaborate situation, an experimentalist may infer 
if an expected noise characteristic is in agreement with the measured syndrome data.

In summary, the Steane code serves as an example that illustrates how one can obtain information about a noise profile of 
a communication channel. In a QKD setting, the code's syndrome distribution may be used to expose such unwanted disturbances. We expect that this approach can be fruitful in a regime where the channel noise strongly exceeds circuit-level noise; even though the distribution may still exhibit structure in the presence of strong circuit-level noise. An appropriate metric to measure how closely two given syndrome distributions align would be desirable. We conjecture that an intentional design of QEC codes to serve the detection and analysis of specific noise maps in the QKD pipeline is possible, for instance, based on QEC code properties such as soundness and confinement that describe relations between errors and syndromes~\cite{campbell2019theory}. It would be interesting to further investigate the behavior of logical qubits in an attack scenario.


\section{Conclusion}\label{sec:conclusion}

In this work, we present the QI-Nutshell framework, which enables the emulation of quantum communication procedures on an ion-trap quantum processor by mapping the communication protocols to sequences of gate operations. Quantum information is transferred within the employed quantum processor by physically shuttling ions in space, providing an illustrative analogy to quantum communication protocols based on photons transferred through fiber-based or free-space quantum communication channels. Therefore, our approach offers an accessible means to experimentally investigate quantum communication protocols that are within and beyond the reach of current quantum communication hardware.

\subsection{Technical summary}

We implement the BB84 QKD protocol and perform a PCCM attack as well as an attack using an imbalanced cloner. By introducing deterministic and probabilistic errors throughout the circuit, we demonstrate the versatility and flexibility of our approach. Notably, without any hardware adaptations on the trapped-ion architecture, we furthermore realize the BBM92 protocol based on the creation and distribution of entangled pairs of qubits.

A QML approach is employed to learn parameters of a PCCM attack. For specified correlations between the measurement outcomes of Alice, Bob, and Eve, an optimized attack angle is found. Although the observed correlations deviate from the specified values due to the noise present in the hardware, a noiseless simulation of the PCCM attack using the experimentally identified optimal attack angle exhibits correlations that closely align with the target value. This means that the optimization procedure reliably learns the attack angle corresponding to the minimum of the cost function also in the presence of noise.

Additionally, we demonstrate side-channel attack mechanisms that are usually hard to include in formal security proofs of QKD protocols but can compromise the communication channel between Alice and Bob. Specifically, we demonstrate the acquisition of information about and the manipulation of Bob’s measurement outcome. The presented tools are parameterized so that a trade-off between the eavesdropper's information gain and its discoverability can be emulated. The demonstrated attack mechanisms can be combined in a single execution of the emulation, enabling the investigation of intricate side-channel attack strategies.

Furthermore, we investigate the integration of QED and QEC codes in emulated QKD protocols. Numerical simulations employing various noise models for the quantum communication channel suggest that the implementation of the $[[4,2,2]]$ QED code can provide a reduction of the QBER. Experimental realizations of the $[[4,2,2]]$ code in various quantum computing platforms demonstrate its practicability in NISQ era devices~\cite{linke2017fault, takita2017experimental, erhard2021entangling}. 

We also find that QEC codes, apart from reducing the QBER, can be used to infer noise characteristics of the channel. The stabilizer generator expectation values of the $[[7,1,3]]$ code are utilized to exemplarily distinguish noise processes that predominantly affect a specific qubit or apply a certain Pauli error with a higher probability. As the stabilizer measurements are sensitive to certain types of errors, a monitoring of the quantum channel’s noise properties and the detection of potential eavesdropping becomes feasible. This diagnostic capability of QEC codes in QKD is shown to remain intact even in a regime where the QEC implementation itself is noisy.

\subsection{Future of QI-Nutshell}

QI-Nutshell is developed for prototyping and testing quantum communication protocols. The approach allows to study the impact of emulated components of the communication processes. As the emulation of these components directly integrates the associated quantum interaction processes, there is an advantage compared to a simulator.

First and foremost, we can integrate and analyze the effects of disturbances in a quantum communication protocol. Basic noise models that are found in well-known software libraries, like \texttt{qiskit}, \texttt{pennylane} or \texttt{tket}, may be implemented directly at the quantum level of the protocol. Moreover, more sophisticated noise profiles with space and time dependence could be realized in our hardware platform. 

QI-Nutshell may turn out as a useful tool to advance our current understanding of quantum communication protocols in the presence of realistic perturbations that cannot easily be treated analytically. It is an open question and thus needs further analysis, whether novel quantitative findings about actual physical realizations of quantum communication protocols can be deduced from QI-Nutshell emulations.

Second, QI-Nutshell is highly accessible. It is possible to instruct the platform to emulate a given communication protocol, without the necessity to develop specific theoretical models including all quantum processes. This opens the door for stakeholders from related fields, for instance cybersecurity, to build their intuition and skills for the quantum age. Easily accessible tools, that take interdisciplinary aspects into account are important for the development of meaningful quantum applications and use cases.

Note that these advantages can already be harnessed with current small-scale devices. We therefore believe that QI-Nutshell is a NISQ-era use case for quantum computers. Any practical use for early-stage hardware is highly desirable, especially if it allows for enabling non-experts to use and experience quantum technologies in a meaningful way. However, more research and development is needed to unlock the full potential of QI-Nutshell. First, we are aiming to increase the faithfulness of QI-Nutshell in regard to the emulation of real implementations of QKD. This involves the integration of realistic adapted noise models and various side-channels, e.g.~photon loss and detector efficiency mismatch, along the lines of Refs.~\cite{trefilov2024intensitycorrelationsdecoystatebb84, sixto2025quantumkeydistributionimperfectly, nahar2025imperfectdetectorsadversarialtasks}. To this end, it is desirable to define an emulation score to quantify the reliability of QI-Nutshell compared to idealized simulations. Second, we aim to deepen the understanding of the channel monitoring tool introduced in this work. Third, we want to develop and test further quantum cryptographic protocols, possibly also involving the distribution of data instead of random key material including but not limited to quantum secure direct communication~\cite{QSDC}. This line of research, for which QI-Nutshell may represent a suitable testbed, has recently gained attention~\cite{bhattacharyya2025uncloneableencryptiondecoupling,yamaguchi2025encryptedqubitscloned, goswami2025hybridauthenticationprotocolsadvanced}.

Ultimately, we aim to connect the fields of quantum communication, quantum internet, 
quantum networks, quantum computing and cryptography among stakeholders from research, industry and governance in order to help realizing new applications from interdisciplinary research.

\section*{Code availability}
Software code used in this project is available from the corresponding authors upon reasonable request.

\section*{Author contributions}
JH ran the experiments. JH, SH and WW performed the QKD simulations. SH performed the QEC simulations. JH, SH, AG, AW, FSK, LP, UP and WW wrote the manuscript with feedback from all authors.

\section*{Acknowledgments}

We thank the Bundesdruckerei-Innovation leadership and team for their support and encouragement. We also thank the team at JoS QUANTUM for their feedback.

\bibliographystyle{apsrev4-2_modified}
\bibliography{lit}

\end{document}